\documentclass[fleqn,twoside]{article}
\usepackage{amsmath,amssymb,espcrc2,color}

\usepackage{graphicx,soul}
\usepackage[figuresright]{rotating}

\newcommand{\url}[1]{{\tt #1}}
\newcommand{\lsim}
{\;\raisebox{-.3em}{$\stackrel{\displaystyle <}{\sim}$}\;}

\newcommand{\gmt}{\ensuremath{(g-2)_\mu}}

\newcommand{\bsg}{BR($b \to s \gamma$)}

\newcommand{\bmm}{BR($B_s \to \mu^+\mu^-$)}
\newcommand{\ssi}{\sigma^{\rm SI}_p}
\newcommand{\Och}{\ensuremath{\Omega_\chi h^2}}

\newcommand{\Mh}{M_h}

\newcommand{\mt}{m_t}
\newcommand{\mgl}{m_{\tilde g}}

\newcommand{\neu}[1]{\tilde \chi^0_{#1}}
\newcommand{\mneu}[1]{m_{\tilde \chi^0_{#1}}}

\newcommand{\mstaue}{m_{\staue}}
\newcommand{\staue}{\tilde \tau_1}
\newcommand{\tb}{\tan\beta}

\newcommand{\tev}{\,\, \mathrm{TeV}}
\newcommand{\gev}{\,\, \mathrm{GeV}}
\newcommand{\mev}{\,\, \mathrm{MeV}}

\definecolor{Orange}{named}{Orange}
\definecolor{Purple}{named}{Purple}

\graphicspath{{figs/}}

\title{\bf Frequentist Analysis of the Parameter Space of Minimal Supergravity
  \\ \vspace{0.5em}}

\author{
{\bf O.~Buchmueller}\address[Imperial]
   {High\,Energy\,Physics\,Group, Blackett\,Laboratory, Imperial\,College, 
    Prince\,Consort\,Road, London\,SW7\,2AZ,\,UK},
{\bf R.~Cavanaugh}\address[FNAL]
   {Fermi National Accelerator Laboratory, P.O. Box 500, 
    Batavia, Illinois 60510, USA}\hbox{$^{\rm ,}$}\address[UIC]
   {Physics Department, University of Illinois at Chicago, Chicago, 
    Illinois 60607-7059, USA},
{\bf D.~Colling}\addressmark[Imperial],
{\bf A.~De Roeck}\address[CERN]
   {CERN, CH--1211 Gen\`eve 23, Switzerland}\hbox{$^{\rm ,}$}\address[Antwerpen]
   {Antwerp University, B--2610 Wilrijk, Belgium},
 {\bf M.J.~Dolan}\address[IPPP]
   {Institute for Particle Physics
     Phenomenology,\,University\,of\,Durham,\,South 
     Road,\,Durham\,DH1\,3LE,\,UK},
{\bf J.R.~Ellis}\addressmark[CERN]\hbox{$^{\rm ,}$}\address[KCL]{Theoretical Physics
  and Cosmology Group, Department of Physics, King's College London, London
  WC2R 2LS, UK}, 
{\bf H.~Fl\"acher}\address[Rochester]
   {Department of Physics and Astronomy, University of Rochester, 
    Rochester, New York 14627, USA},
{\bf S.~Heinemeyer}\address[Santander]
   {Instituto de F\'{\i}sica de Cantabria (CSIC-UC), 
    E--39005 Santander, Spain},
{\bf K.A.~Olive}\address[Minnesota] 
   {William\,I.\,Fine\,Theoretical\,Physics\,Institute,\,University\,of\,Minnesota,\,Minneapolis,\,Minnesota\,55455,\,USA},  
{\bf S.~Rogerson}\addressmark[Imperial],
{\bf F.J.~Ronga}\address[ETHZ]
   {Institute for Particle Physics, ETH Z\"urich, CH--8093 Z\"urich, 
   Switzerland},
{\bf G.~Weiglein}\address[DESY]
   {DESY, Notkestrasse 85, D--22607 Hamburg, Germany}
}
       
\begin{document}

\begin{abstract}
We make a frequentist analysis of the parameter space of
minimal supergravity (mSUGRA), in which, as well as the gaugino and scalar
soft supersymmetry-breaking parameters being universal,
there is a specific relation between the trilinear, bilinear and scalar
supersymmetry-breaking parameters, $A_0 = B_0 + m_0$, and the
gravitino mass is fixed by 
$m_{3/2} = m_0$. We also consider a more general model,
in which the gravitino mass constraint is relaxed (the VCMSSM).
We combine in the global likelihood function the experimental constraints
from low-energy electroweak precision data, the anomalous
magnetic moment of the muon, the lightest Higgs boson mass~$\Mh$, 
$B$~physics and the
astrophysical cold dark matter density, assuming that the lightest
supersymmetric particle (LSP) is a neutralino. In the VCMSSM, we find a
preference for values of 
$m_{1/2}$ and $m_0$ similar to those found previously in frequentist
analyses of the constrained MSSM (CMSSM) and a model
with common non-universal Higgs masses (NUHM1).
On the other hand, in mSUGRA we find two preferred regions: one with
larger values of both $m_{1/2}$ and $m_0$ than in the VCMSSM, and one
with large $m_0$ but small $m_{1/2}$. 
We compare the probabilities of the
frequentist fits in mSUGRA, the VCMSSM, the CMSSM and the NUHM1:
the probability that mSUGRA is consistent with
the present data is significantly less than in the other models.
We also discuss the mSUGRA and VCMSSM predictions for sparticle masses
and other observables, identifying potential signatures at the LHC and
elsewhere. 

\bigskip
\begin{flushleft}
\vspace{-0.5cm}
\end{flushleft}
\begin{center}
{\tt CERN-PH-TH/2010-169, DCPT-10-196, DESY 10-211, IPPP-10-98,
  FTPI-MINN-10/34, KCL-PH-TH/2010-34, UMN-TH-2928/10, arXiv:1011.6118 [hep-ph]}
\end{center}
\vspace{-0.5cm}
\end{abstract}

\maketitle

\section{Introduction}
\label{sec:intro}

One of the most favoured possible extensions of the Standard Model (SM)
is supersymmetry (SUSY), which renders natural the electroweak mass
scale \cite{hierarchy} and accommodates 
grand unification of the particle interactions \cite{GUT}. If $R$~parity
is conserved 
it also provides a promising candidate for astrophysical cold dark matter,
which might be the lightest neutralino,
$\neu{1}$ \cite{EHNOS}, or the gravitino
\cite{EHNOS,gdm1,gdm2}. SUSY also predicts the appearance of 
a relatively light Higgs boson \cite{erz,Degrassi:2002fi}, and may
provide a welcome correction to the SM prediction for the anomalous
magnetic moment of the muon, \gmt~\cite{newBNL,g-2,newDavier}.

However, even the minimal supersymmetric extension of the SM, 
the MSSM \cite{MSSM}, 
boasts over 100 free parameters, mostly associated with
the mechanism of soft SUSY-breaking. Hence simplified scenarios with
particular restrictions on the pattern of SUSY-breaking are often
studied. One example is the constrained MSSM (CMSSM) 
\cite{cmssm1,cmssm2,like1,Ellis:2006ix}, in which the
soft SUSY-breaking gaugino masses $m_{1/2}$, scalar masses
$m_0$ and trilinear couplings $A_0$ are each assumed to be universal
at the grand unification (GUT) scale, and $\tb$ is unconstrained.
This leads to four effectively-free parameters, if the gravitino
is assumed to be sufficiently heavy and/or rare that its cosmological
decays and its mass are irrelevant. Another possibility is to relax the
universality constraint 
for common soft SUSY-breaking contributions to the Higgs
masses, yielding the NUHM1 \cite{nuhm1} 
with five effective parameters in addition to the gravitino mass
that is assumed to be irrelevant.  

Alternatively, additional assumptions may be imposed, as in minimal
supergravity (mSUGRA)~\cite{Polonyi,Fetal,bfs,CAN,BIM}, 
in which there is a specific relation between the 
trilinear and bilinear soft SUSY-breaking parameters and the
universal scalar mass: $A_0 = B_0 + m_0$, and the gravitino mass is
set equal to the common scalar mass before renormalization:
$m_{3/2} = m_0$~\footnote{See, for example, remark (b) following
equation (16) of~\cite{bfs}.}.  
Hence mSUGRA has just 3 free parameters, namely
$m_{1/2}, m_0$ and $A_0$, and $\tb$ is now fixed by the radiative
electroweak symmetry breaking conditions \cite{rewsb}.
Further, there is a  
restriction on $m_0$ if the lightest supersymmetric particle (LSP) is
the lightest neutralino, $\neu{1}$, as we assume here for the reasons
discussed below. An intermediate scenario is the very constrained MSSM
(VCMSSM), in which again $A_0 = B_0 + m_0$ but the gravitino mass is
left free, so there is no restriction 
on $m_0$ \cite{vcmssm}. If $m_{3/2}$ is sufficiently large,
and/or the gravitino abundance is sufficiently low (as we will assume
here), there are no related 
cosmological constraints, and the VCMSSM also has effectively 3 relevant
free parameters, but they are less constrained than in mSUGRA.

We have previously published frequentist analyses of the CMSSM and NUHM1
parameter spaces \cite{mc1,mc2,mc3,mc35,mc-web}, implementing the experimental
constraints from low-energy electroweak precision data and the lightest
Higgs boson mass $\Mh$ as well as the lower limits from the direct
searches for SUSY particles at LEP, 
fitting the measured value of the anomalous 
magnetic moment of the muon, \gmt, $B$~physics and the
cosmological dark matter density, \Och, assuming that the
LSP is the lightest neutralino, $\neu{1}$. In this paper
we extend these analyses to include the VCMSSM and mSUGRA which have,
as discussed above, one or two additional constraints on the
pattern of soft SUSY-breaking, respectively. 
An early $\chi^2$ analysis in these scenarios can be found
in~\cite{Ellis:2006ix}. 

In the case of the VCMSSM, in which $A_0 = B_0 + m_0$ but the gravitino mass is
free, we assume that the gravitino is sufficiently heavy or rare that the dark
matter is composed of neutralinos and the cosmological effects of its
decays are unimportant. Under these assumptions, 
as we show below, imposing the neutralino dark matter constraint does not
increase substantially the $\chi^2$ of the global minimum, which is
$\sim 1.2$ higher than in the CMSSM, but
we find 68 and 95\% confidence-level (CL) ranges of $m_{1/2}$ and $m_0$
that are more restrictive than those found previously in our analyses of
the CMSSM and the NUHM1.

However, the preferred part of the VCMSSM parameter space has
$\mneu{1} > m_0$, which within mSUGRA would imply that 
$m_{3/2} = m_0 < \mneu{1}$, so that the lightest neutralino would be
unstable, and the dark matter would be composed of gravitinos. 
In such a case, the usual calculation of the neutralino dark matter
density \Och\ would be inapplicable, and one should, instead, consider the
constraints on the decays of long-lived neutralinos or other sparticles
into gravitinos that are imposed by the cosmological abundances of light
elements \cite{gdm2,cefos,gdecay}.  
In fact, these constraints are sufficiently strong to exclude, within
mSUGRA and assuming a standard
cosmological evolution, the otherwise preferred regions of parameter space
with $m_0 = m_{3/2} < \mneu{1}$ and hence gravitino dark matter
\cite{cefos}~\footnote{See~\cite{gg} for possible effects of
  non-standard cosmological histories that might invalidate these
  arguments against the gravitino LSP scenario.}. 
Thus, the surviving region of the mSUGRA parameter space
has $m_{3/2} = m_0 > \mneu{1}$, corresponding to 
neutralino dark matter. 

We find within the mSUGRA model two distinct regions with local minima
of the $\chi^2$ function, each of which is significantly worse than in the
VCMSSM, namely $\Delta \chi^2 \sim 7$ or $11$ with 19 dof,
corresponding to a goodness--of--fit of just 6.0\% 
(or 2.3\%) in the mSUGRA hypothesis.
This may be contrasted with the cases of the CMSSM and the NUHM1 studied in
\cite{mc1,mc2,mc3,mc35}, where the best-fit parameters are
consistent with the current experimental constraints at the level of
32\% (31\%) fit probability, and with the VCMSSM case (20 dof, 
31\% fit probability). 
One of the regions preferred in mSUGRA has larger
values of both 
$m_{1/2}$ and $m_0$ than in the VCMSSM, CMSSM and NUHM1,
and the other has larger $m_0$ but small $m_{1/2}$.


\section{Notations}
\label{sec:notations}

Before describing our analyses of the VCMSSM and mSUGRA in more detail, 
we first specify our notations, 
since different conventions for the MSSM superpotential couplings and 
the trilinear and bilinear soft SUSY-breaking
terms are used elsewhere in the literature, including in~\cite{mc1}. 
Our conventions here follow those specified, e.g., 
in~\cite{dlt}, according to which the superpotential includes the terms
\begin{align}
W &\ni Y_e \epsilon H_1 L E^c + Y_d \epsilon H_1 Q D^c \\
&\quad    + Y_u \epsilon H_2 Q U^c + \mu \epsilon H_1 H_2~, \nonumber
\end{align}
where the $Y_i$ are Yukawa couplings, $\epsilon$ is the antisymmetric 
$2 \times 2$ tensor with $\epsilon_{12} = + 1$, $H_{1,2}, L, E^c, Q, D^c$ 
and $U^c$ are superfields, and $\mu$ is the Higgs supermultiplet mixing
parameter. The corresponding trilinear and 
bilinear SUSY-breaking terms in the effective Lagrangian are:
\begin{equation}
{\cal L} \; \ni \; - (A_t Y_t \epsilon h_2 q t^c + \dots)
                   - \mu B \epsilon h_1 h_2,
\label{ABterms}
\end{equation}
where the lower-case letters denote the scalar components of the
corresponding superfields. Within this convention, 
$\sin 2 \beta = - 2 B \mu/(m_1^2 + m_2^2 + 2 \mu^2)$ at the tree level, where
$m_{1,2}$ is the soft SUSY-breaking mass of $H_{1,2}$, the
left-right mixing term in the stop mixing matrix is 
$m_{LR}^2 = - m_t (A_t + \mu \cot \beta)$,
and the one-loop renormalization of the trilinear coefficient has the form
$d A_t / dt \ni - \frac{16}{3} g_3^2 M_3 + 6 Y_t^2 A_t + \dots$
(where $M_3$ denotes the soft SUSY-breaking parameter in the gluino
  sector, and $A_t$ is the trilinear Higgs-stop coupling.)
These choices unambiguously determine the sign conventions for $A$ and $B$.

Within this convention, $A_0 = B_0 + m_0$  before renormalization in mSUGRA
with its minimal (flat) K\"ahler potential~\footnote{Note that many of
the publicly available packages such as 
{\tt SoftSusy}~\cite{Allanach:2001kg} use the opposite sign convention,
as may be ascertained by comparing the signs of the gauge and Yukawa
contributions to the RGEs of the $A$ parameters.  In the notation of
these codes, the mSUGRA boundary condition would be $A_0 = B_0 - m_0$.}. 
Additionally, as already mentioned, the choice
of a minimal K\"ahler potential also imposes the condition $m_{3/2} =
m_0$ on the gravitino mass before renormalization, so that mSUGRA 
has just three independent parameters.


\section{Details of the evaluation}
\label{sec:details}

Our analysis has been performed using the
{\tt MasterCode}~\cite{mc1,mc2,mc3,mc35,mc-web}. 
We sample the VCMSSM and mSUGRA parameter spaces using a
Markov Chain Monte Carlo (MCMC) technique similar
to that used in our previous analyses of the CMSSM and
NUHM1~\cite{mc1,mc2,mc3,mc35}. We evaluate the global likelihood using a
$\chi^2$ function 
constructed by combining the likelihoods for the experimental constraints
from electroweak precision data, the anomalous
magnetic moment of the muon, \gmt, $B$ physics, the
astrophysical cold dark matter density, \Och, and searches for the
lightest Higgs boson and supersymmetric particles, in exactly the same way as
described previously~\cite{mc1,mc2,mc3}. The most 
significant change in our numerical treatments of these
observables since~\cite{mc3} is in \gmt, for which we use use the estimate
$a_\mu^{\rm SUSY} = (28.7 \pm 8.0) \times 10^{-10}$~\cite{newDavier}.

The numerical evaluation within the {\tt MasterCode}
combines the following theoretical codes. For the RGE running of
the soft SUSY-breaking parameters, it uses
{\tt SoftSUSY}~\cite{Allanach:2001kg}, which is combined consistently
with the codes used for the low-energy observables: 
{\tt FeynHiggs}~\cite{Degrassi:2002fi,Heinemeyer:1998np,Heinemeyer:1998yj,Frank:2006yh}  
is used for the evaluation of the Higgs masses and
$a_\mu^{\rm SUSY}$  (see also
\cite{Moroi:1995yh,Degrassi:1998es,Heinemeyer:2003dq,Heinemeyer:2004yq}).
For flavour-related observables we use 
{\tt SuFla}~\cite{Isidori:2006pk,Isidori:2007jw} and 
{\tt SuperIso}~\cite{Mahmoudi:2008tp,Eriksson:2008cx}, and
for the electroweak precision data we have included 
a code based on~\cite{Heinemeyer:2006px,Heinemeyer:2007bw}.
Finally, for dark-matter-related observables, we use
{\tt MicrOMEGAs}~\cite{Belanger:2006is,Belanger:2001fz,Belanger:2004yn} and
{\tt DarkSUSY}~\cite{Gondolo:2005we,Gondolo:2004sc}. 
We make extensive use of the SUSY Les Houches
Accord~\cite{Skands:2003cj,Allanach:2008qq} 
in the combination of the various codes within the {\tt MasterCode}.

Our MCMC sampling of the VCMSSM parameter space comprises some 30,000,000
points. The neutralino CDM constraint on \Och\ \cite{WMAP} and the Higgs 
mass constraint~\cite{Barate:2003sz,Schael:2006cr} were applied after the
sampling, allowing the effects of these two constraints to be studied
separately. In the case of mSUGRA, about 17,000,000 of the MCMC points
from the VCMSSM sample survive the mSUGRA constraint $m_0 = m_{3/2} >
\mneu{1}$, and we again applied the neutralino 
CDM constraint on \Och\ and the Higgs mass constraint {\it a posteriori}. 


\section{Analysis of parameter planes}
\label{sec:results}

\begin{figure*}[htb!]
\resizebox{8cm}{!}{\includegraphics{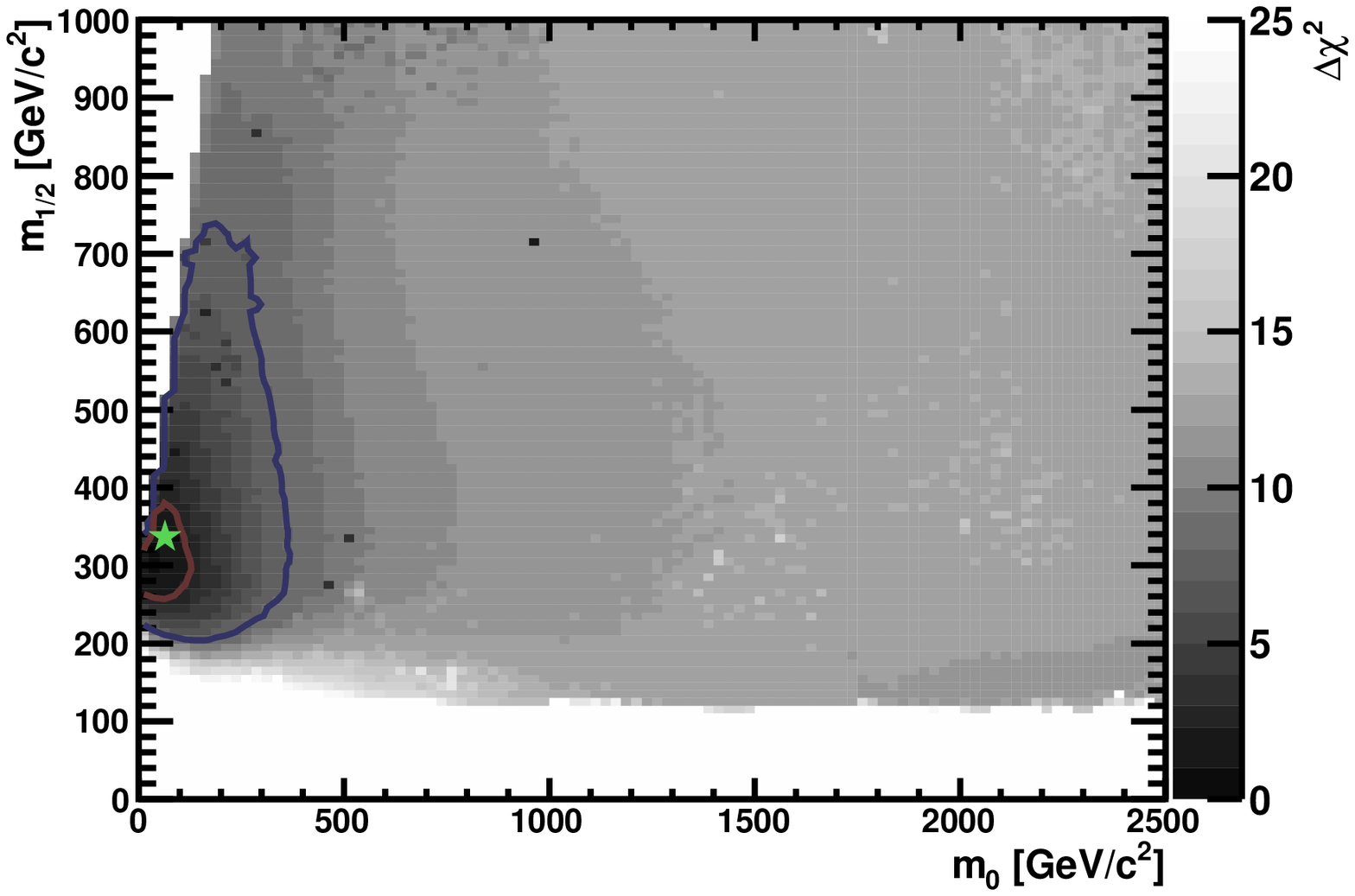}}
\resizebox{8cm}{!}{\includegraphics{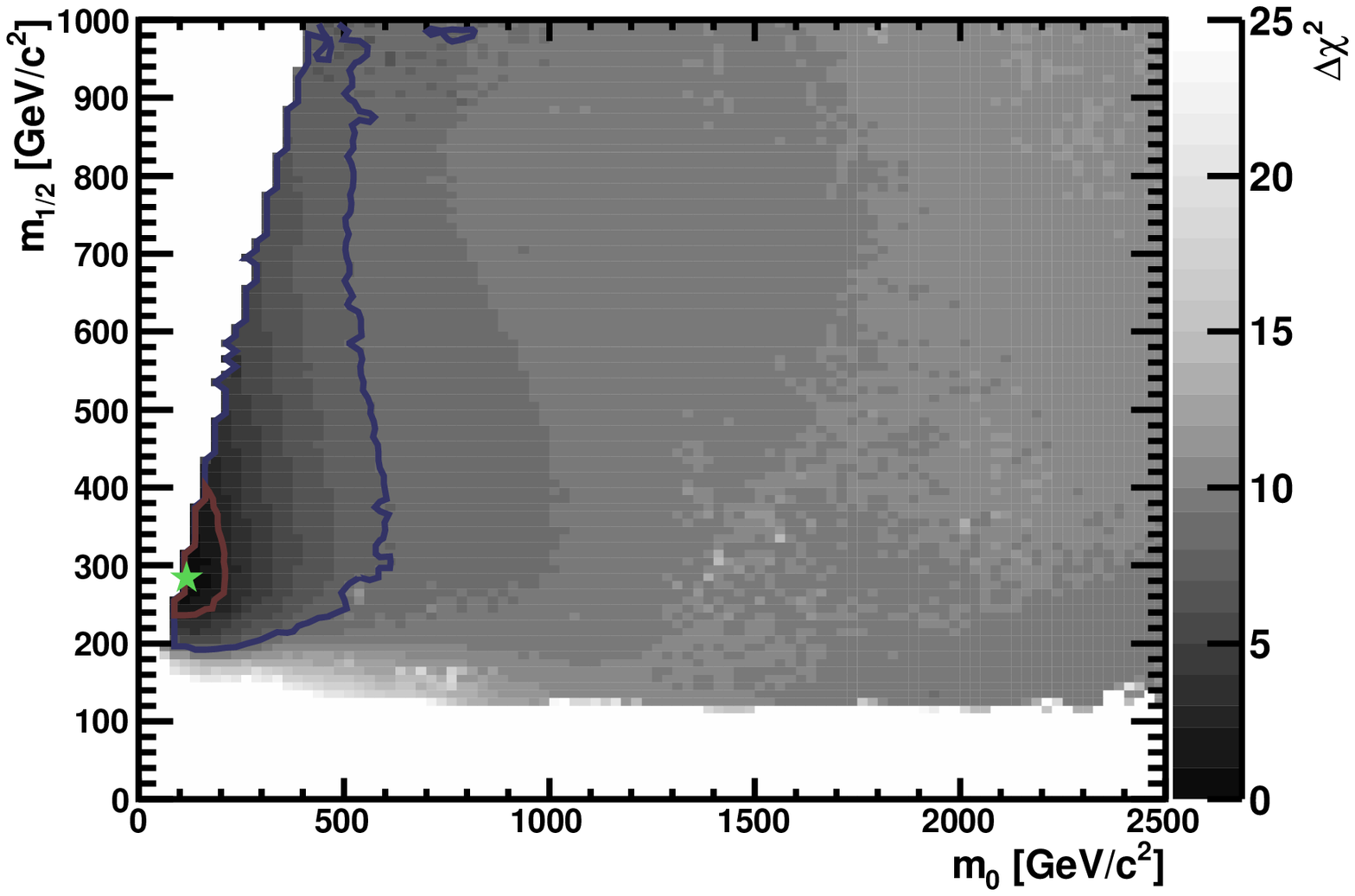}}
\resizebox{8cm}{!}{\includegraphics{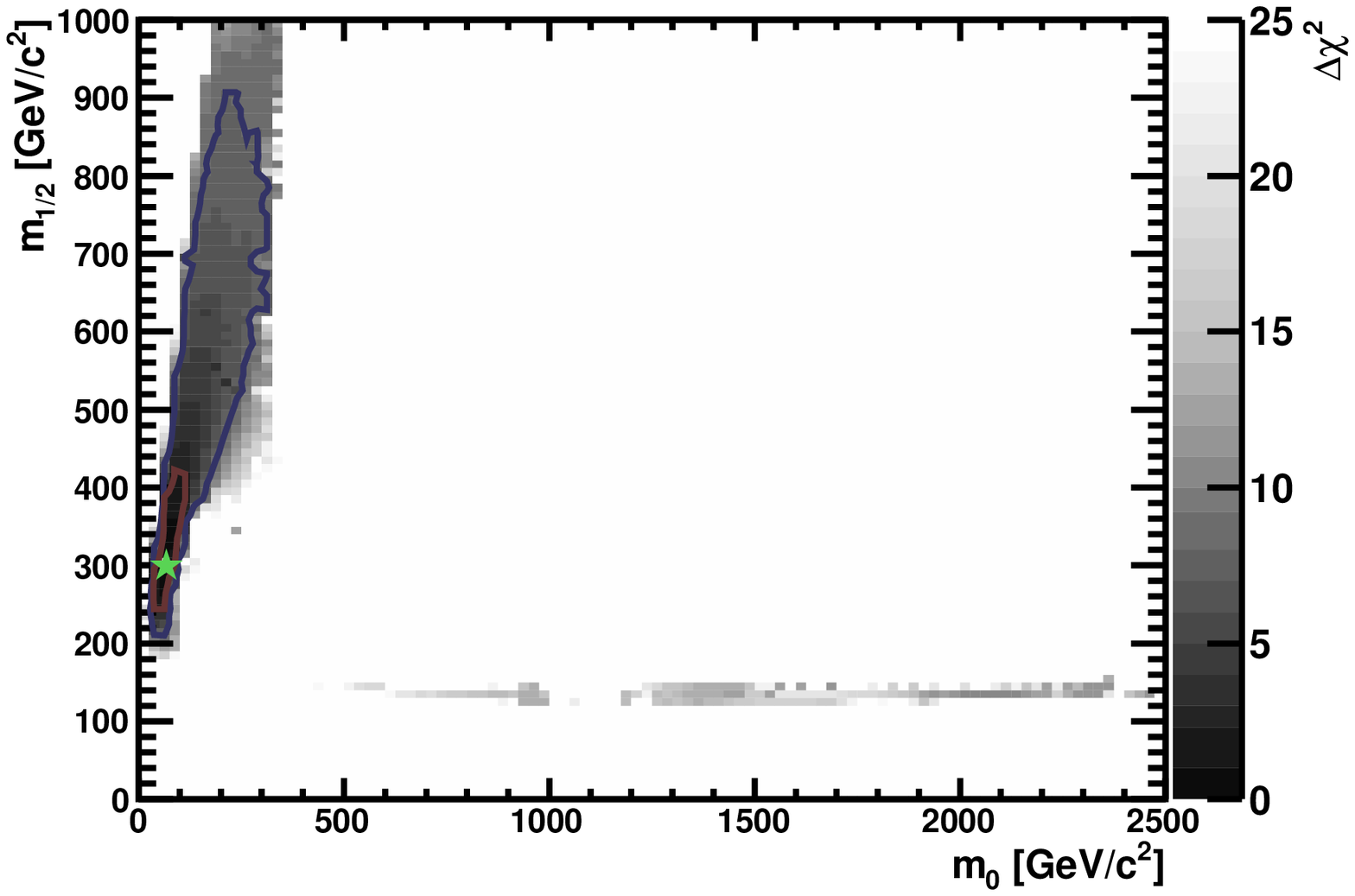}}
\resizebox{8cm}{!}{\includegraphics{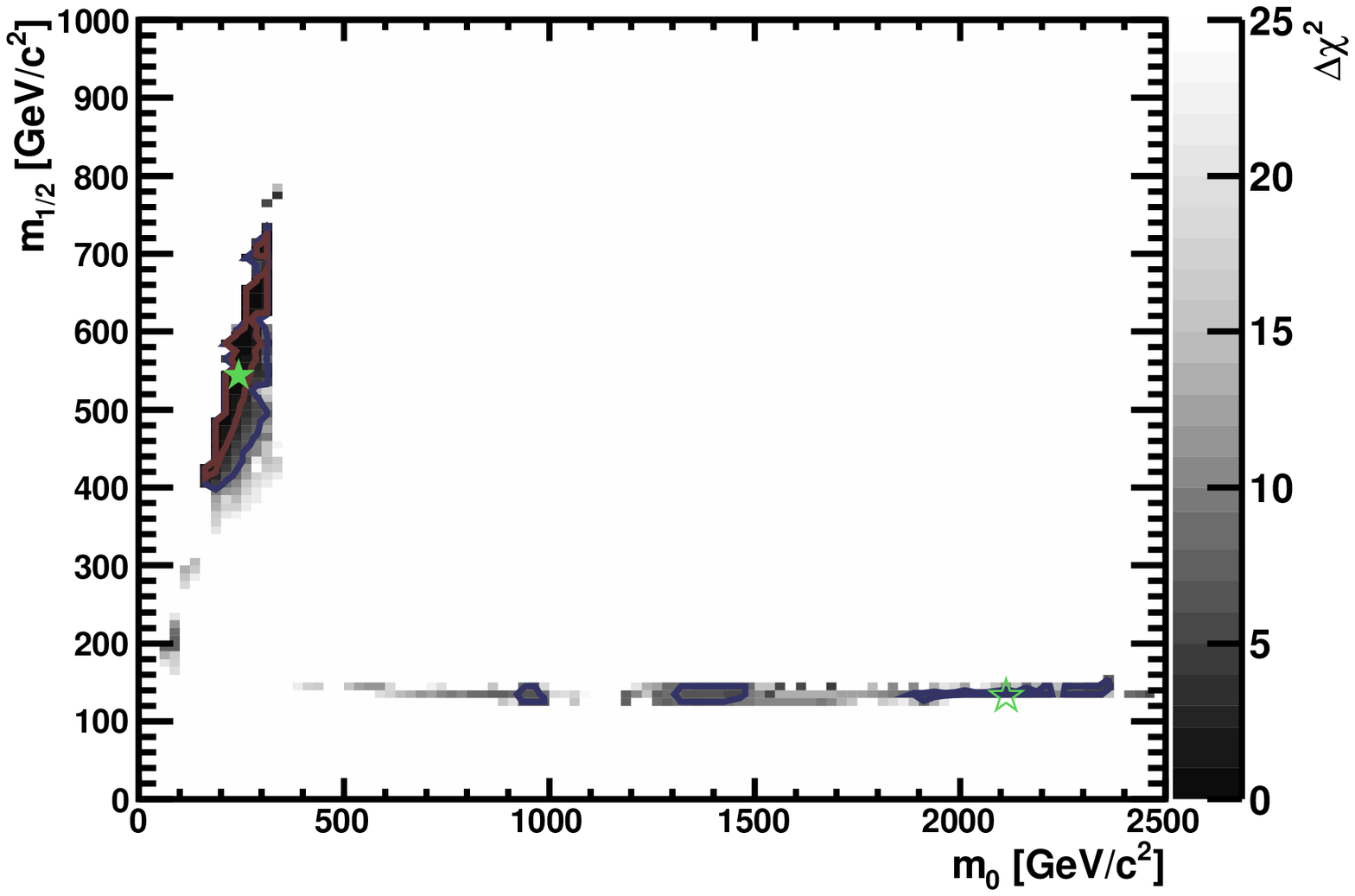}}\\
\vspace{-1cm}
\caption{\it The $(m_0, m_{1/2})$ planes in the VCMSSM (left panels) and
  mSUGRA (right panels), without (upper) and with (lower panels) the
  \Och\ constraint~\protect\cite{WMAP},  showing in each case the
  best-fit points (green stars) and the 68 and 95\% CL contours 
(red and black, respectively). The open green star in the lower right
  panel denotes the secondary minimum discussed in the text.
}
\label{fig:m0m12}
\end{figure*}

We start our analysis with the results in the $(m_0, m_{1/2})$ planes. 
Fig.~\ref{fig:m0m12} displays the global likelihood functions
in the VCMSSM (left panels) and mSUGRA (right panels). In each
case, the upper panel shows results before the \Och\ constraint is
applied, and the lower panel displays the effects of imposing the \Och\
constraint~\footnote{We recall that the constraints due to the late decays of
massive metastable particles~\cite{cefos} are taken into 
account implicitly as described above, i.e., in the VCMSSM by assuming that the 
gravitino mass is high and/or its primordial density is low,
and by accepting that the decay constraints are so severe in mSUGRA as
to forbid $m_{3/2} = m_0 < \mneu{1}$ in that case.}.
In all panels, we display the points with the minimal values of $\chi^2$
(green stars)
as well as the 68 and 95\%~CL contours (red and blue),
corresponding to $\Delta \chi^2 = 2.28$ and 5.99. Other contours of 
$\Delta \chi^2$ are indicated in shades of grey. 

In the upper left panel for the VCMSSM before applying the \Och\
constraint, the triangular region at small $m_0$ and large $m_{1/2}$ 
is excluded because there the LSP would be charged, and a band
extending to large $m_0$ at low $m_{1/2}$ is excluded by the LEP Higgs
constraint. The best-fit point is at $(m_0, m_{1/2}) = (30, 310) \gev$,
the 68 and 95\% CL contours 
enclose regions of the $(m_0, m_{1/2})$ planes that are similar to those
favoured in the CMSSM and NUHM1~\cite{mc2,mc3}, 
and there are no preferred points in
the focus-point region at large $m_0$. Specifically, we find that the
focus-point region  at large 
$m_0 \sim 2000 \gev$ with $m_{1/2} \sim 200 \gev$ is
subject to a penalty of $\Delta \chi^2 \sim 10$.

In the upper right panel for mSUGRA without the \Och\ constraint, we see a
similar pattern, but with a larger excluded triangular region at large
$m_{1/2}$ and small $m_0$, as the allowed part of the $(m_0, m_{1/2})$
plane is now restricted to the region where 
$m_{3/2} = m_0 > \mneu{1} \sim 0.4 m_{1/2}$.  
Since the best-fit VCMSSM point seen in the upper left panel of
Fig.~\ref{fig:m0m12} lies in the region that is disallowed in mSUGRA,
there is a new mSUGRA best-fit point on the boundary of the
allowed region, with $(m_0, m_{1/2}) = (110, 280) \gev$. 
The minimum value of $\chi^2$ is higher than in the
VCMSSM model by $\sim 1.5$, 
and the $\Delta \chi^2$ values in other regions of the
$(m_0, m_{1/2})$ plane are correspondingly reduced, leading to the emergence
of `archipelago' of points at $(m_0, m_{1/2}) \sim (700, 1000) \gev$
that are now allowed at the 95\% CL. 
The fact that the mSUGRA best-fit point lies on the
  boundary of the allowed parameter space indicates that the
  restrictions in this model are disfavoured by current experimental data.

We recall that it was shown in~\cite{mc2} that in the CMSSM
the \Och\ constraint~\cite{WMAP} has a relatively modest impact
on the preferred ranges in the $(m_0,m_{1/2})$ plane.
However, imposing the \Och\ constraint has a dramatic effect on the VCMSSM
fit, as we see in the lower left panel of Fig.~\ref{fig:m0m12}. The 
region allowed at the 95\% CL is reduced to a narrow `WMAP strip'
terminating at $(m_0, m_{1/2}) \sim (250, 700) \gev$. We recall that
similar WMAP coannihilation strips appear in the CMSSM for fixed values
of $\tb$ and $A_0$, but move across the $(m_0, m_{1/2})$ plane as 
$\tb$ and $A_0$ are varied, which was why the WMAP strip
structure was invisible in the global likelihood fit to the CMSSM
\cite{mc2}. On the other hand, 
we recall that, in the VCMSSM, $\tb$ is fixed
as a function of $m_0, m_{1/2}$ and $A_0$, and the dependence on $A_0$
is not very strong. As a result of the loss of the freedom to vary $\tb$
independently, the WMAP strip structure is resurrected in the VCMSSM.

\begin{figure*}[htb!]
\resizebox{8cm}{!}{\includegraphics{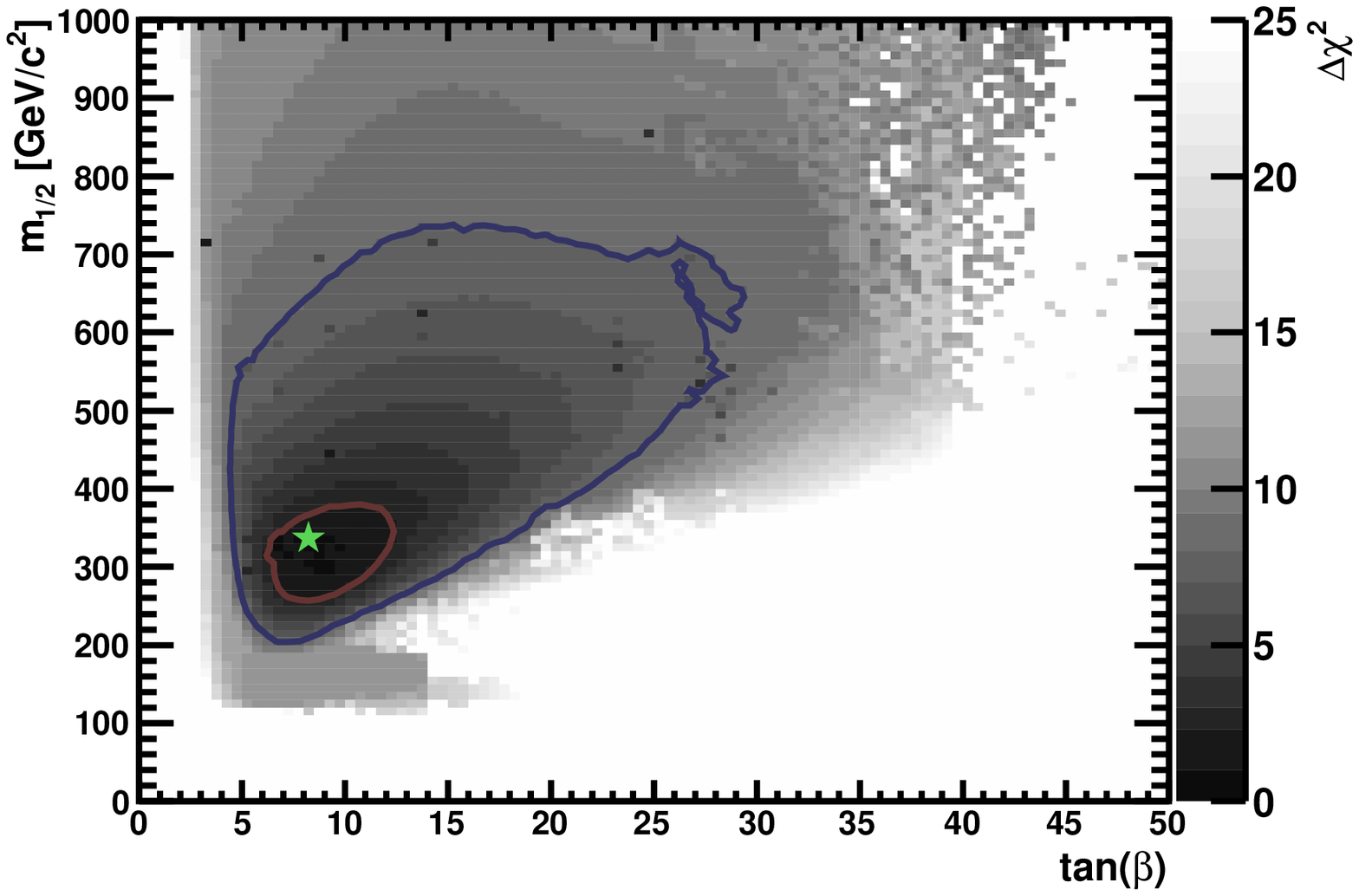}}
\resizebox{8cm}{!}{\includegraphics{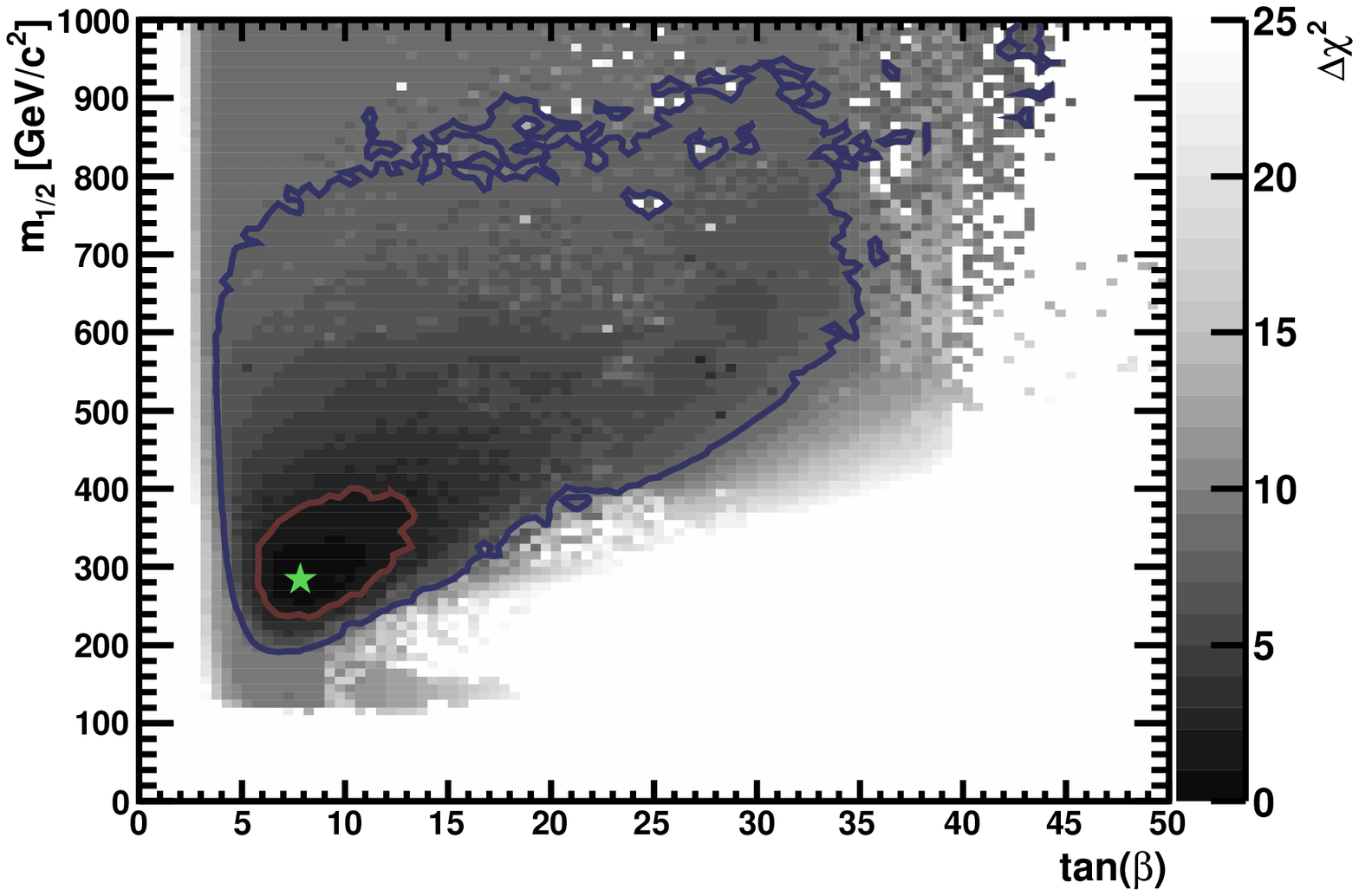}}
\resizebox{8cm}{!}{\includegraphics{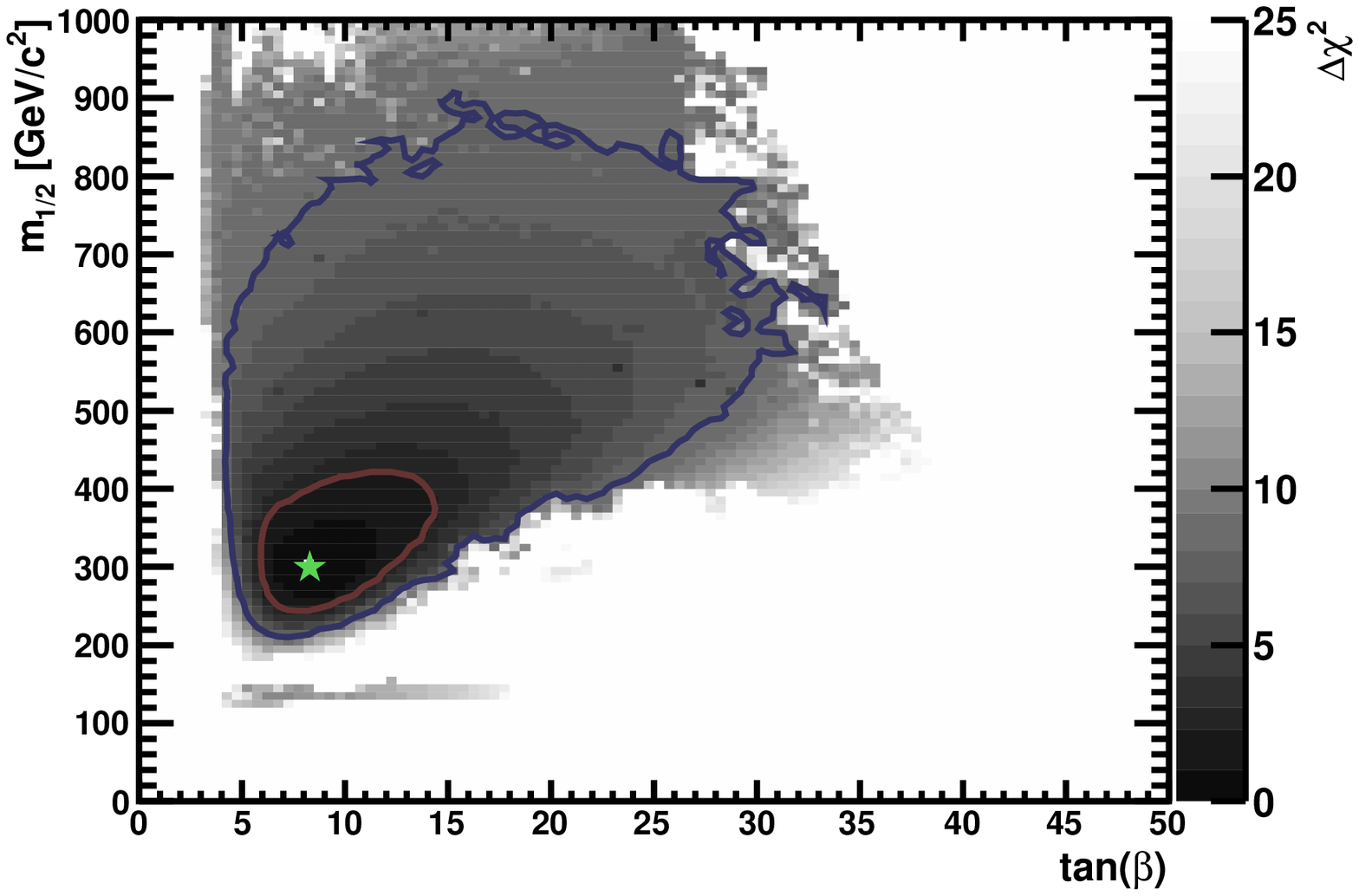}}
\resizebox{8cm}{!}{\includegraphics{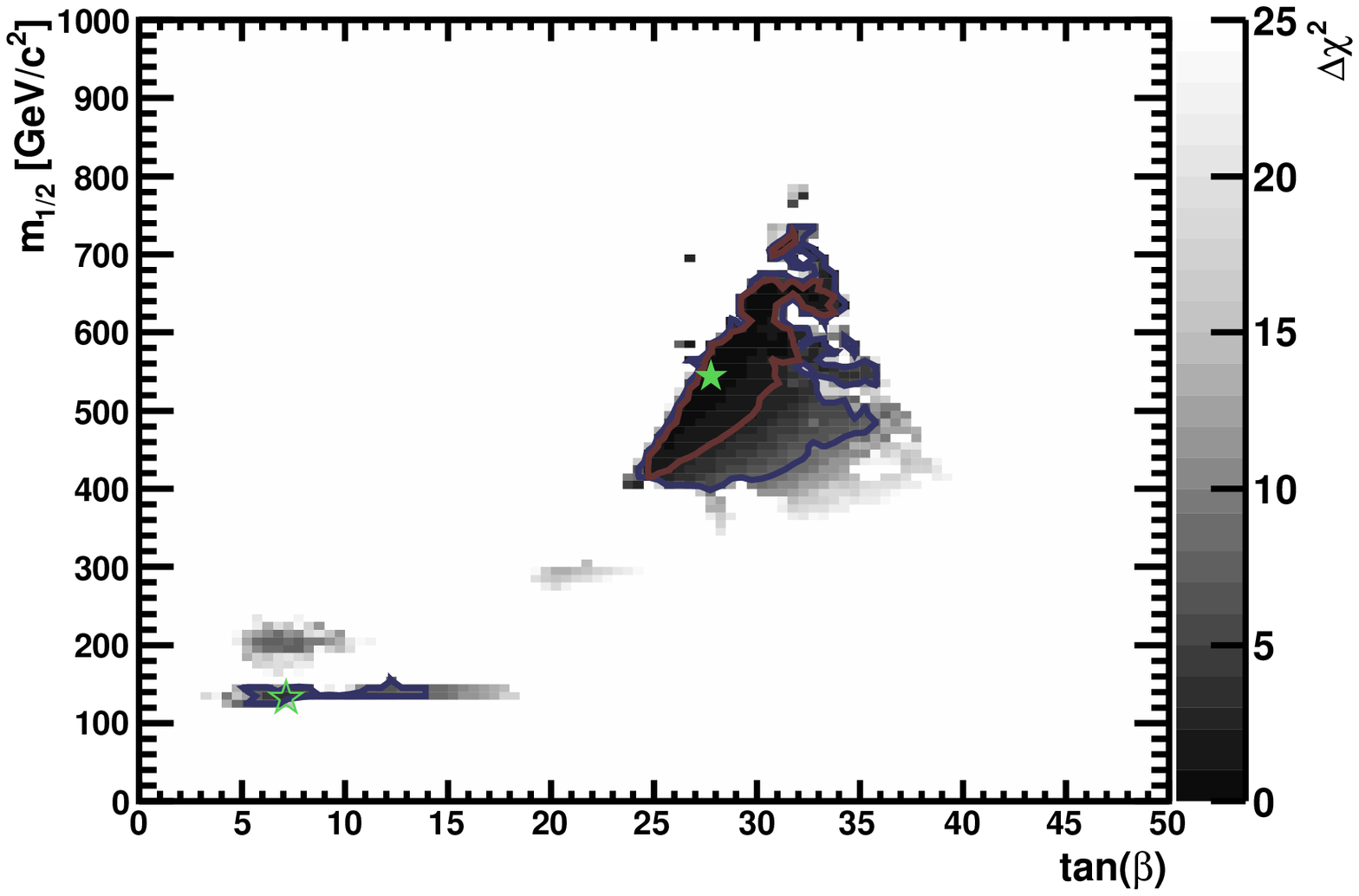}}
\vspace{-1cm}
\caption{\it The $(\tb, m_{1/2})$ planes in the VCMSSM (left) and
  mSUGRA (right), without (upper) and with (lower panels) the
  \Och\ constraint~\protect\cite{WMAP}, showing in each case the
  best-fit points and the 68 and 95\% CL contours. 
}
\label{fig:m12tanb}
\end{figure*}

The best-fit point in the VCMSSM with the \Och\ constraint has
$(m_0, m_{1/2}, A_0) \sim (60, 300, 30) \gev$ and $\tb \sim 9$.
However, as in the previous CMSSM and NUHM1
cases, rather larger values of $m_{1/2}$, and hence 
$\mneu{1}, m_{\tilde g}$ and 
other sparticle masses, are allowed at the 95\% CL. 
This VCMSSM 
fit has a very good value of $\chi^2$/dof = $22.5/20$ 
(31\% probability),
similar to the $\chi^2$ before applying the \Och\ constraint,
demonstrating that there is  
no significant tension between this and other constraints. The
increase in $\chi^2$ due to inclusion of the LEP $\Mh$ 
constraint~\cite{Barate:2003sz,Schael:2006cr} is $\sim 1.1$,
demonstrating that there is also no significant tension
between the $\Mh$ and other constraints. Moreover, the
fact that the fit probability in the VCMSSM is about the same as
the value of 32\% ($\chi^2 = 21.3/19$ dof)
found in the CMSSM indicates that applying
the extra VCMSSM constraint $A_0 = B_0 + m_0$ is certainly not a
source of significant tension in the fit.

We note that there is no focus-point region visible in the VCMSSM
when the \Och\ constraint is applied,
since it is not compatible with the $\tb$ constraint imposed by the
initial conditions. 
On the other hand, there is a very narrow strip at $m_{1/2} \sim 130 \gev$
extending to large $m_0$, where the relic density is brought into the WMAP
range by rapid annihilation through the direct-channel light Higgs
pole. However, 
this strip has $\Delta \chi^2 \ge 9$.

Turning finally to the lower right panel of Fig.~\ref{fig:m0m12} for
mSUGRA with the \Och\ constraint applied, we see an evolution of the
picture.  Much of the VCMSSM `WMAP strip' has disappeared, as only a vestige
of it has $\mneu{1} < m_0 = m_{3/2}$.  Since the minimum value of $\chi^2$
in the VCMSSM was located in the forbidden region with 
$\mneu{1} > m_0 = m_{3/2}$, the minimum value of $\chi^2$ in the mSUGRA
region is significantly higher, specifically $\chi^2 \sim 29$, a price
$\Delta \chi^2 \sim 7$. The best mSUGRA fit is again on the boundary of
the allowed region. Moreover, as indicated by the open green star in the
lower right panel of  
Fig.~\ref{fig:m0m12}, a second local minimum with $\chi^2 \sim 33$
(which is therefore allowed at the 95\% CL) can be found along
the light Higgs rapid-annihilation strip with 
$m_{1/2} \sim 130 \gev$ and $900 \gev \lsim m_0 \lsim 2500 \gev$.
Along this strip, 
the $\chi^2$ function is relatively insensitive to $m_0$, thanks to
approximate compensation between the contributions from \bsg, \gmt\
and the forward-backward asymmetry of $b$ quarks measured at LEP,
decreasing slightly as $m_0$ increases 
up to $m_0 \lsim 2100 \gev$.

Fig.~\ref{fig:m12tanb} displays the $(\tb, m_{1/2})$ planes in the
VCMSSM (left) and mSUGRA (right, where the cut $\mneu{1} < m_0$ was
applied), both without (upper) and with (lower) the \Och\ constraint.
In the upper left panel, we see that a range of $\tb \sim 6$ to
$12$ --- similar to that favoured in the CMSSM and NUHM1 --- is preferred in
the VCMSSM at the 68\% CL, but with a best-fit value $\tb \sim 9$ that
is rather smaller. A range of slightly larger $\tb \sim 6$ to $13$ is
allowed in mSUGRA at the 68\% CL, with a best-fit 
value of $\tb \sim 8$  before imposing the \Och\ constraint. The 95\% CL
ranges of $\tb$ extend to $\sim 30, 35$ in the VCMSSM and mSUGRA,
respectively.
When the \Och\ constraint is imposed on the VCMSSM (lower left),
the ranges of $\tb$ favoured at the 68 and 95\% CL are little changed, 
with $\tb \sim 9$ at the best-fit point. We
again see at higher $\Delta \chi^2$ the rapid-annihilation Higgs
funnel at $m_{1/2} \sim 130 \gev$, separated from the favoured
coannihilation region. However, when the \Och\ constraint is imposed on
mSUGRA (lower right), the coannihilation region shrinks to a vestigial
region with $\tb \sim 30$ and relatively high $\chi^2$,
as previously remarked, and the other minimum along the light Higgs
funnel has $\Delta \chi^2 \sim 4$ and $\tb \sim 5$ to 14, and also contains an
area allowed at the 95\%~CL.

\begin{figure*}[htb!]
\resizebox{8cm}{!}{\includegraphics{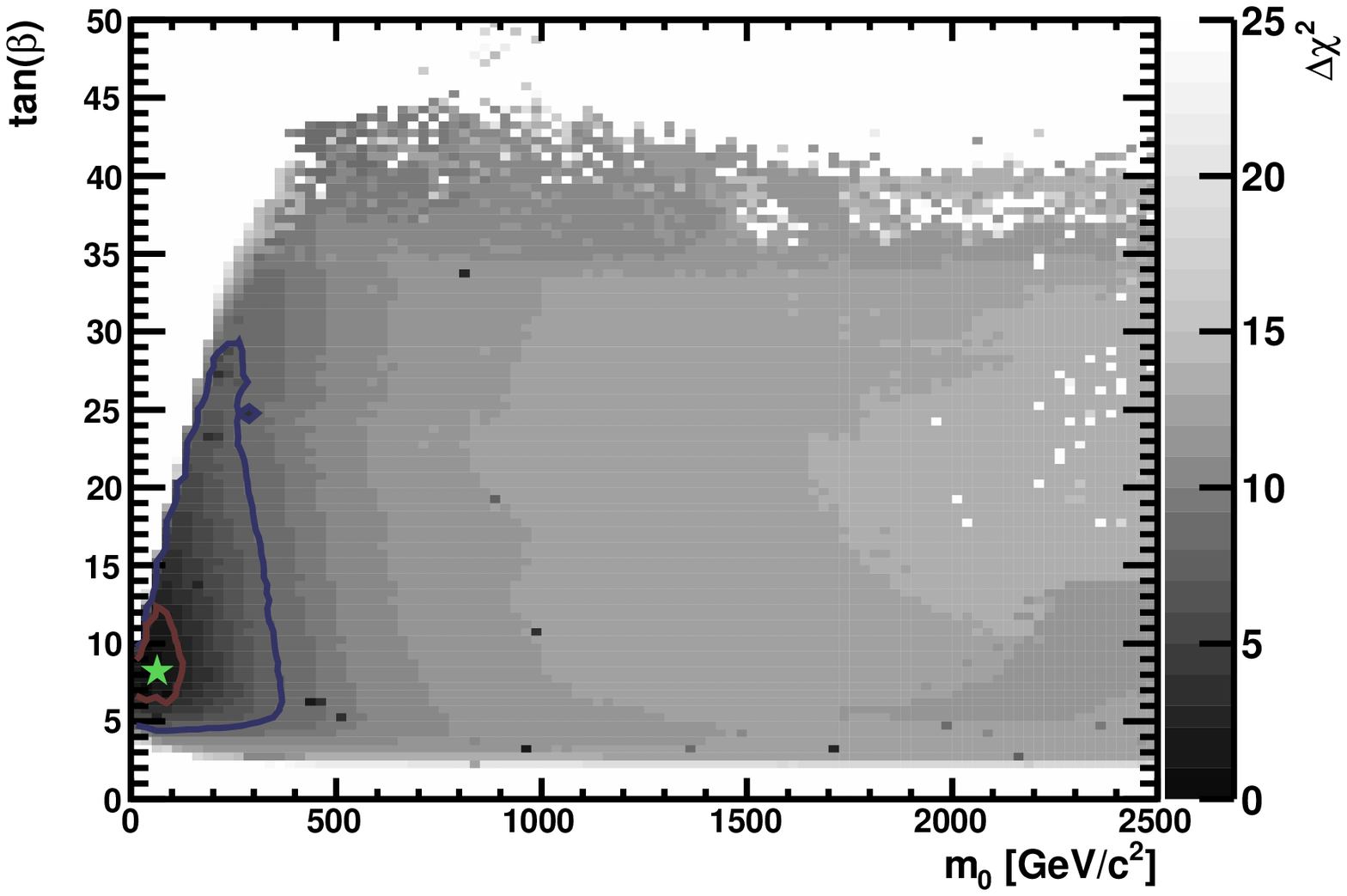}}
\resizebox{8cm}{!}{\includegraphics{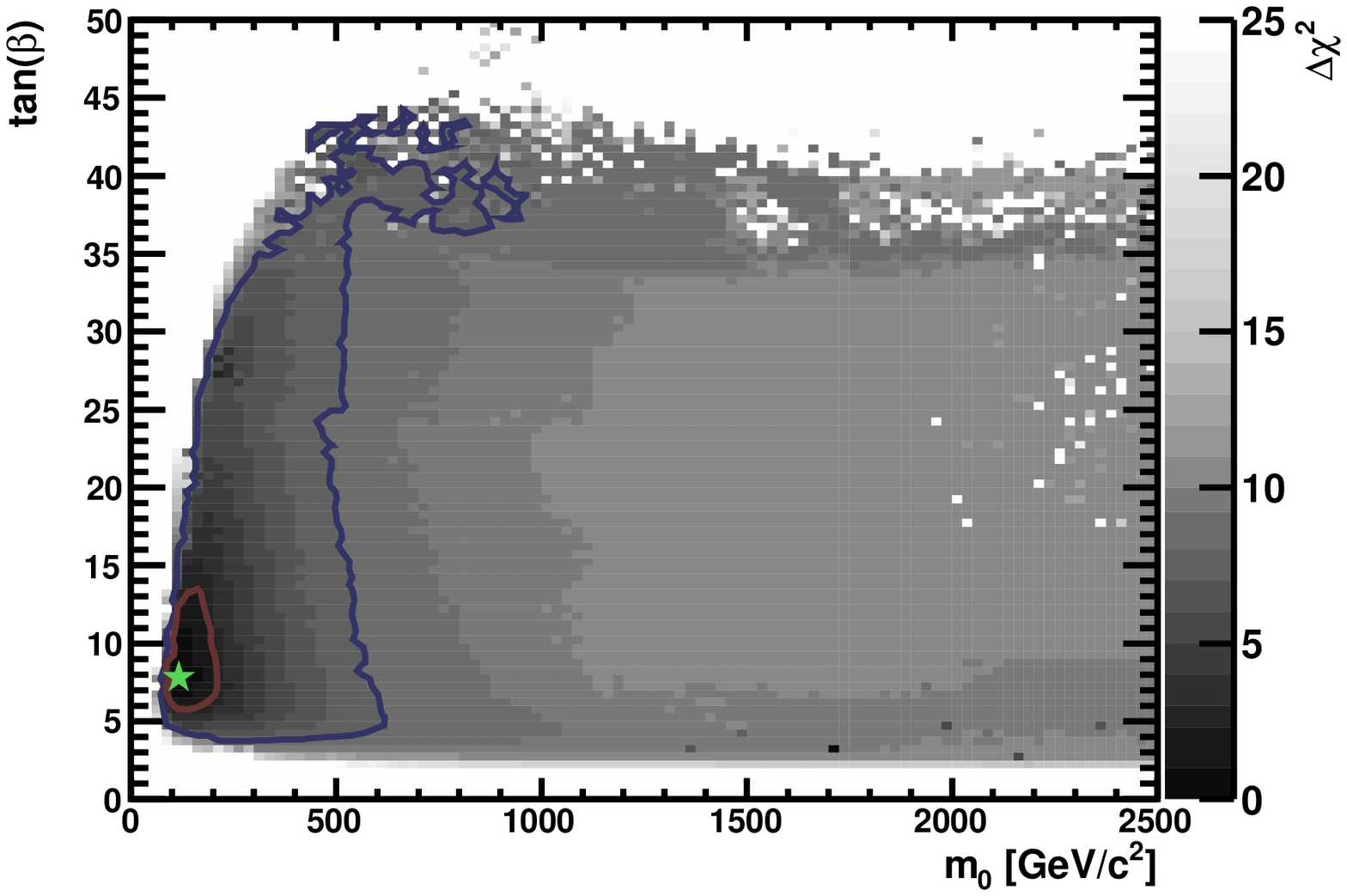}}
\resizebox{8cm}{!}{\includegraphics{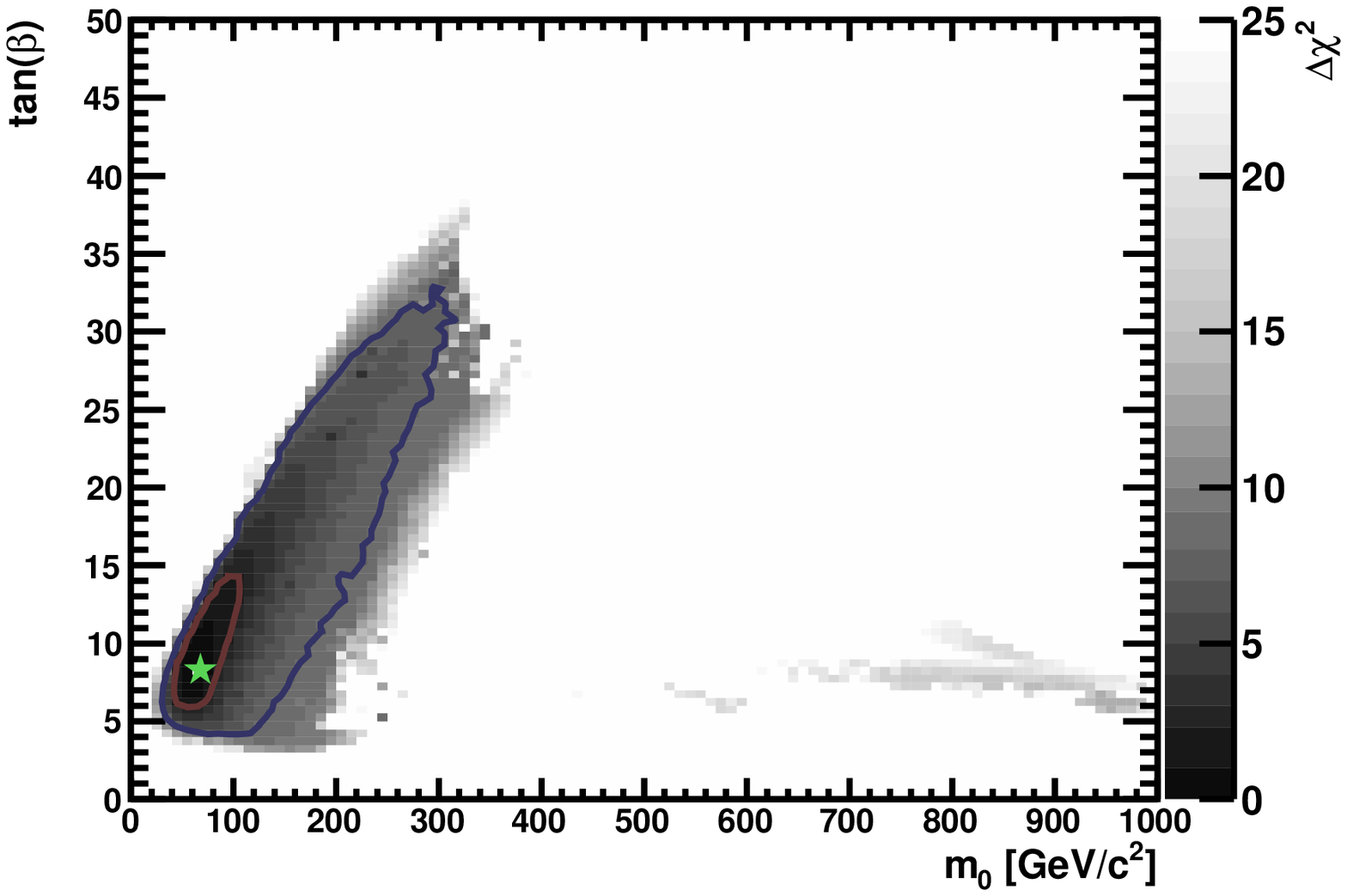}}
\resizebox{8cm}{!}{\includegraphics{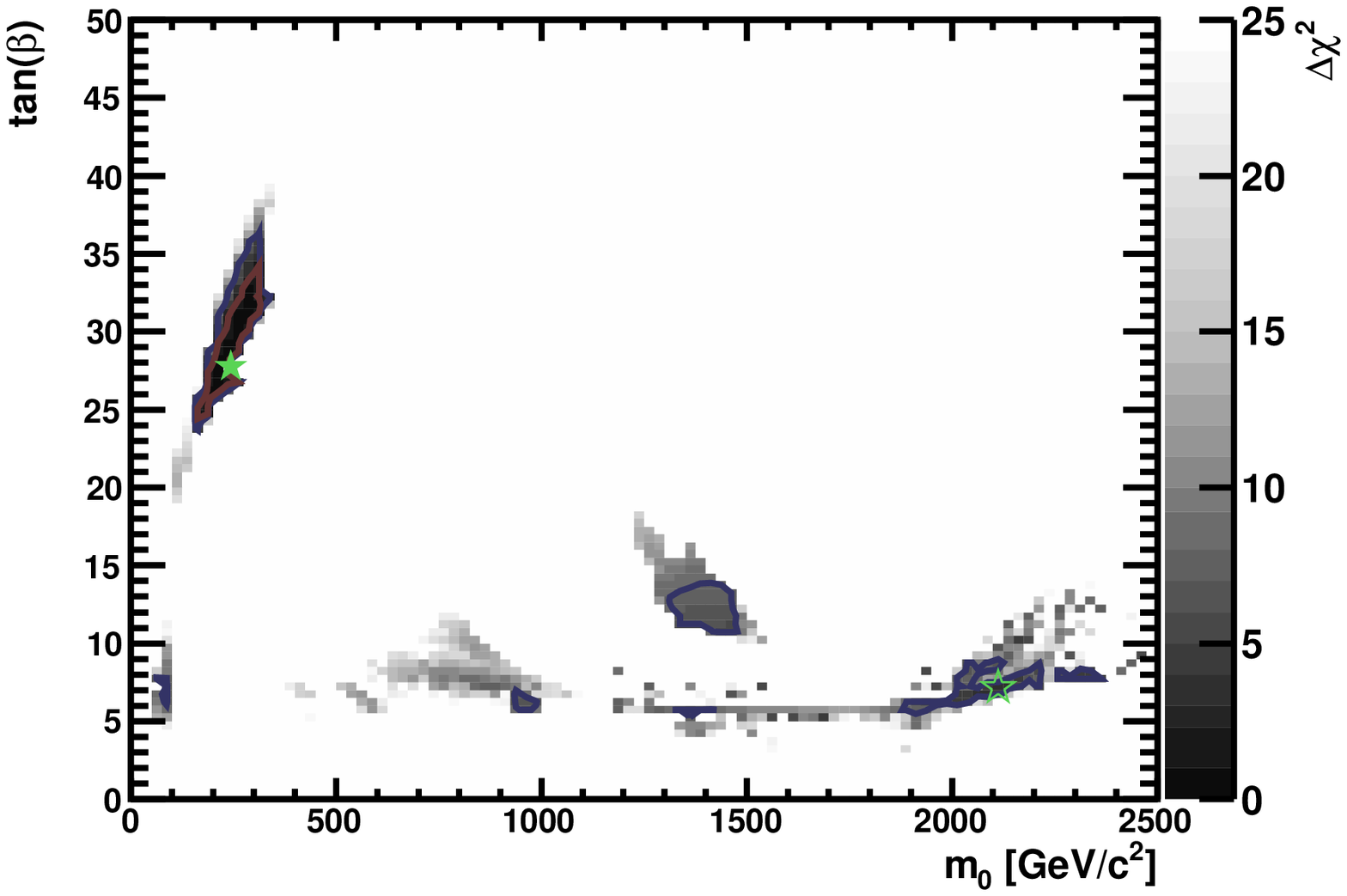}}
\vspace{-1cm}
\caption{\it The $(m_0, \tb)$ planes in the VCMSSM (left) and
  mSUGRA (right), without (upper) and with (lower panels) the
  \Och\ constraint~\protect\cite{WMAP}, showing in each case the
  best-fit points and the 68 and 95\% CL contours. 
}
\label{fig:m0tB}
\end{figure*}

Fig.~\ref{fig:m0tB} displays the corresponding $(m_0, \tb)$ planes
in the VCMSSM (left) and mSUGRA (right). In the absence of the
\Och\ constraint (upper panels), we see regions allowed at the 95\% CL
that are restricted to $m_0 \sim 300 \gev$ in the VCMSSM case (left) 
and $\sim 500 \gev$ ($\sim 1000 \gev$ including the `archipelago')
in the mSUGRA model (right), the larger range being
expected from the restriction $m_0 > \mneu{1}$.
Imposing the \Och\ constraint on the VCMSSM (lower left), we find a band 
at low $m_0$ that can be identified with the WMAP-compatible
coannihilation strip in the corresponding panel of
Fig.~\ref{fig:m0m12}.
Turning to the case of mSUGRA with the 
\Och\ constraint (lower right), only a vestige of the low-$m_0$ band
remains at relatively large $\tb$, and the 
$\chi^2$ in this region is $\sim 4$ smaller than in the
rapid-annihilation region at 
larger $m_0$ and smaller $\tb$ values~\footnote{Because the
rapid-annihilation region is 
very narrow in $m_{1/2}$, and
  quite broad in $m_0$ and $A_0$, it is difficult to sample fully using
  the MCMC approach, even with the high statistics of our full
  sample. Moreover, the MCMC approach samples the input variables
  $m_{1/2}, m_0$ and $A_0$, contributing also to the uneven sampling in
  the derived quantity 
$\tb$ seen in the lower panels of Fig.~\ref{fig:m0tB}.}. 

\begin{figure*}[htb!]
\resizebox{8cm}{!}{\includegraphics{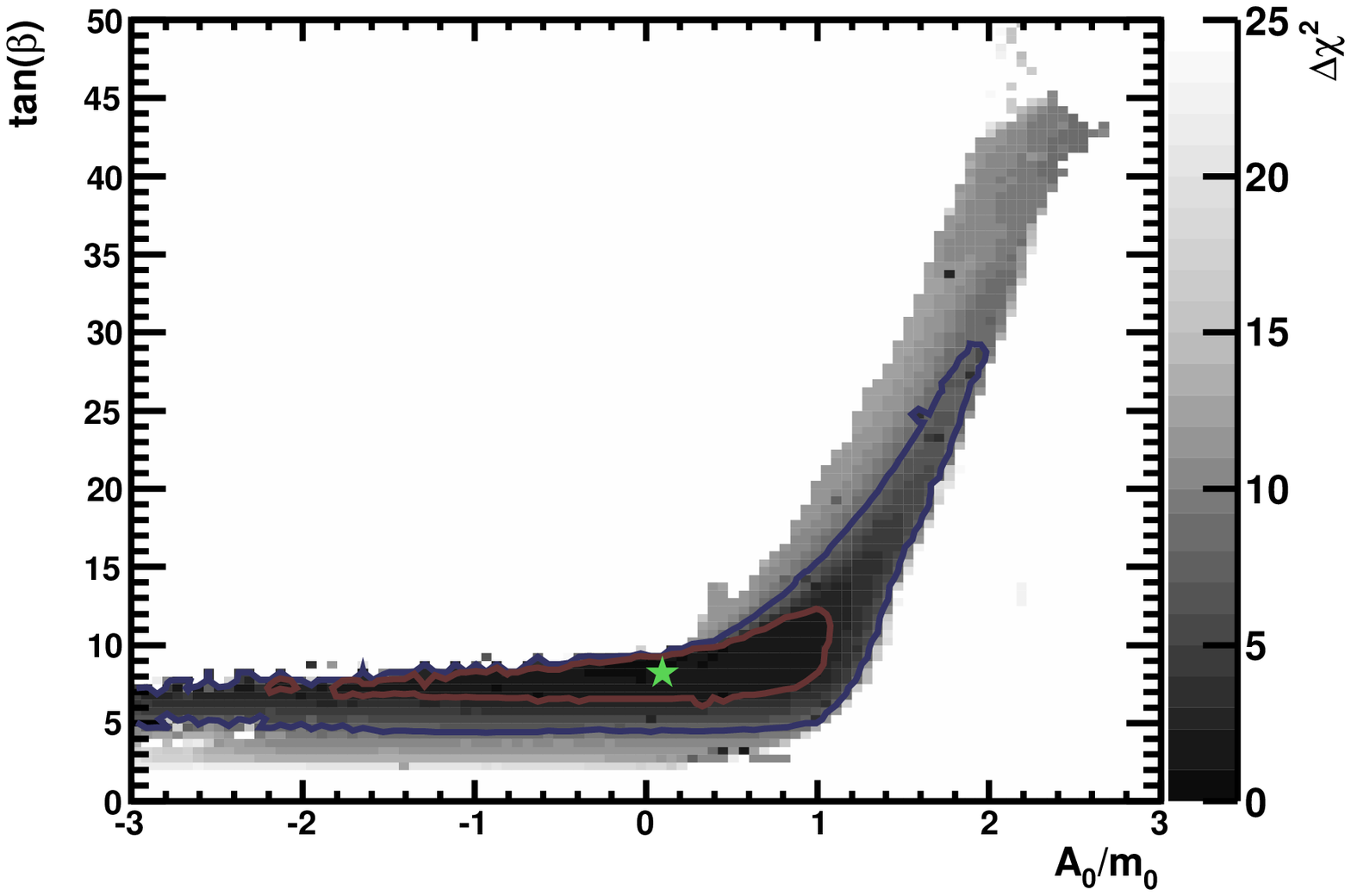}}
\resizebox{8cm}{!}{\includegraphics{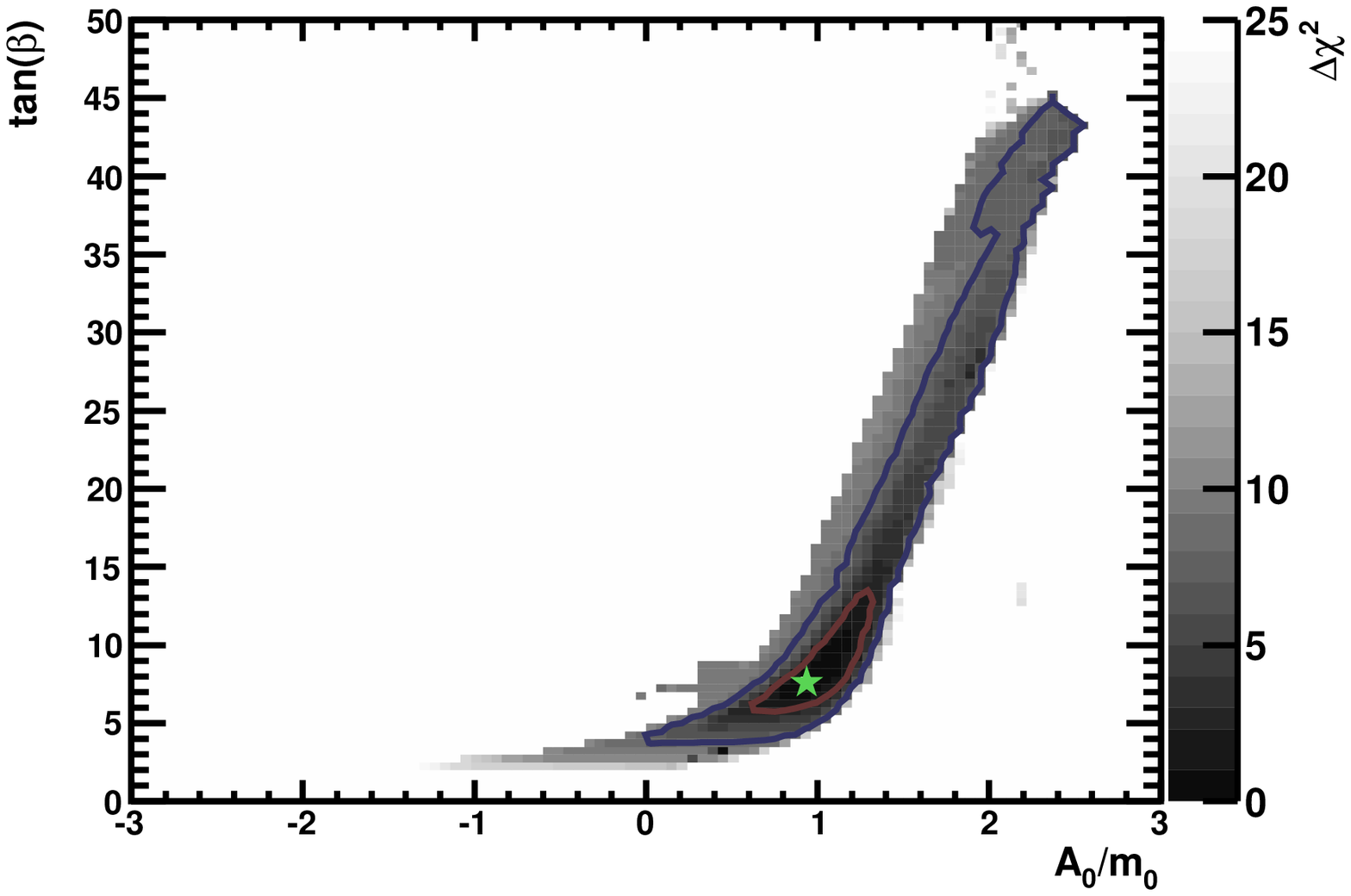}}
\resizebox{8cm}{!}{\includegraphics{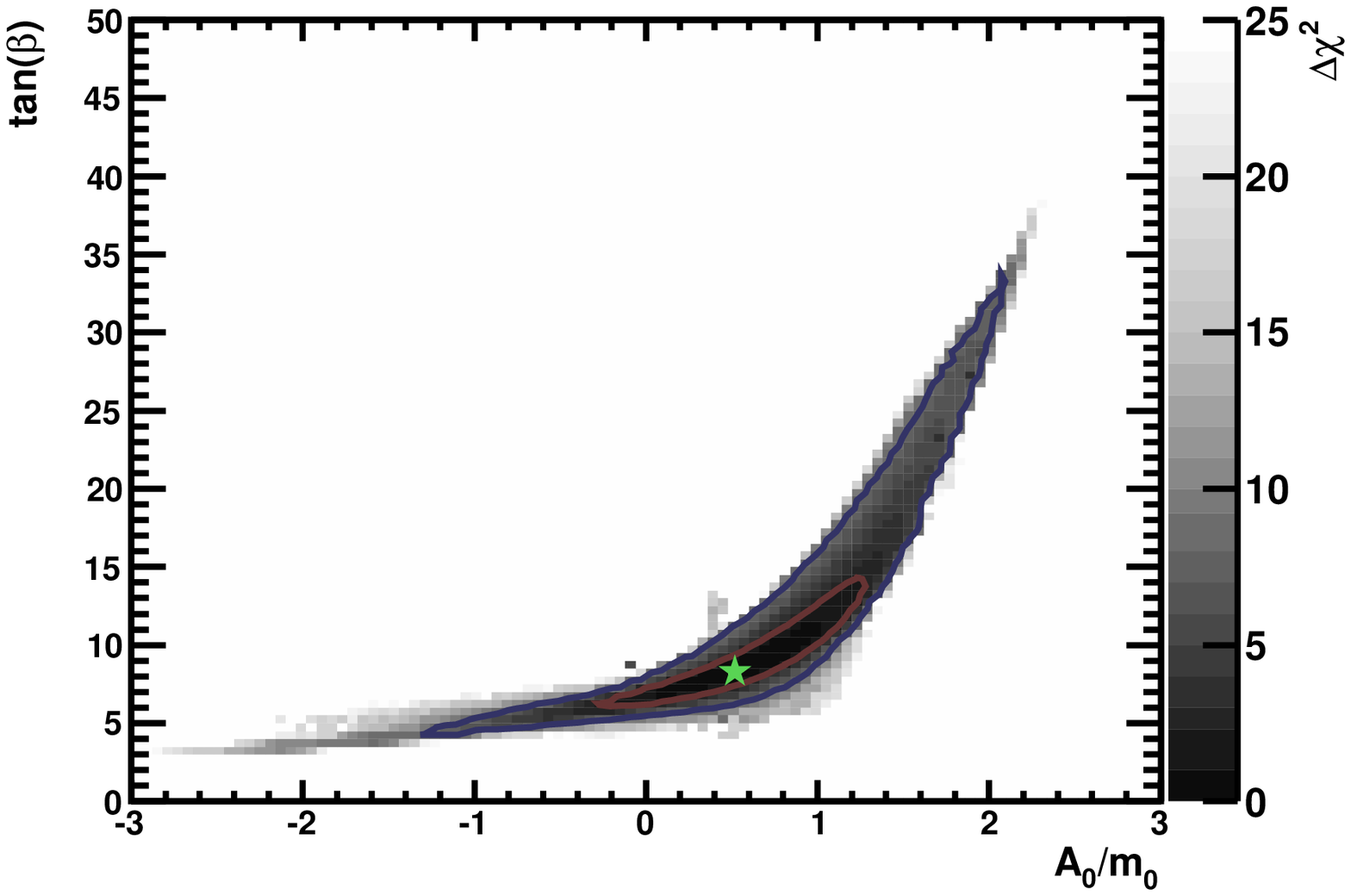}}
\resizebox{8cm}{!}{\includegraphics{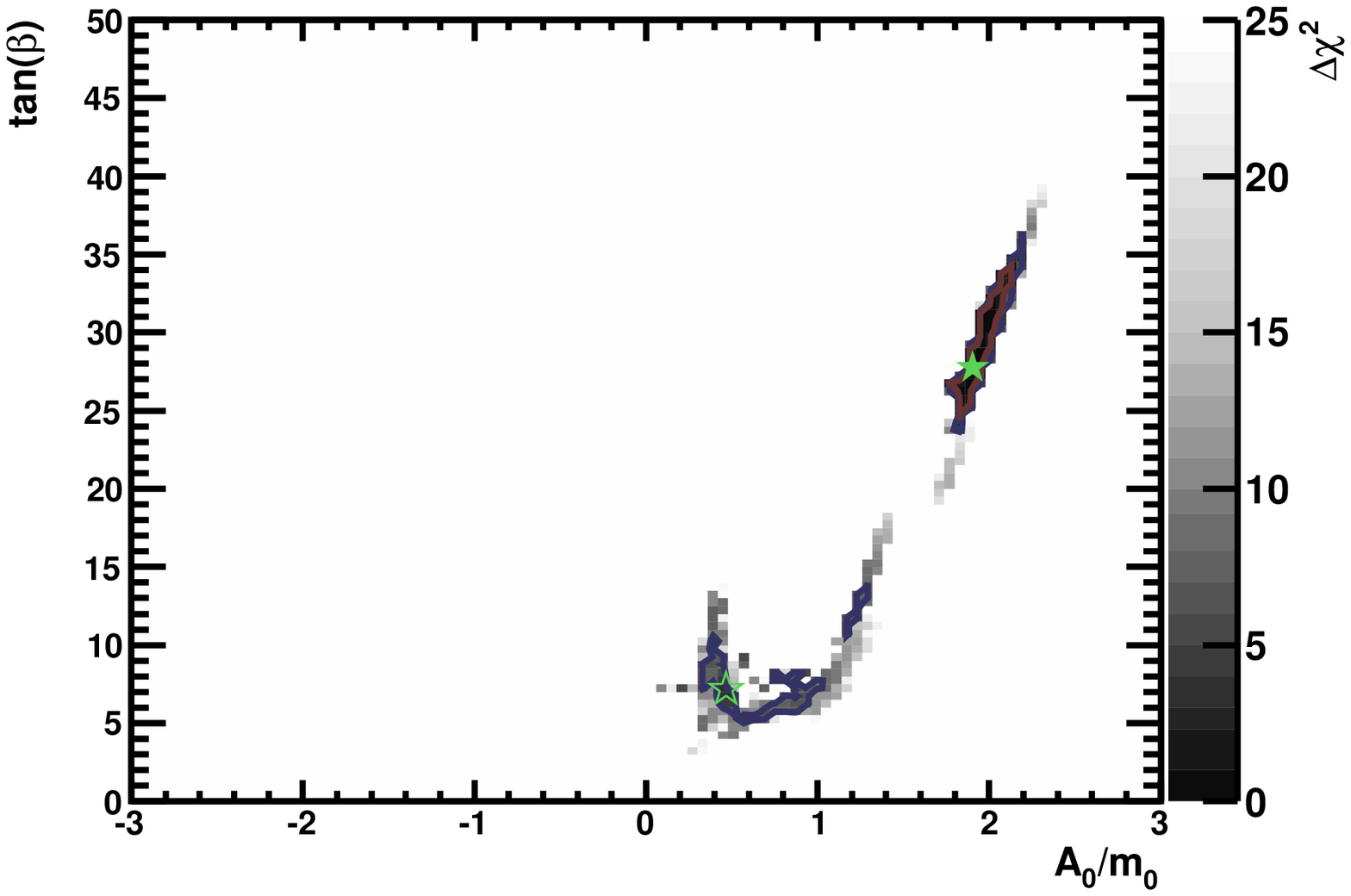}}
\vspace{-1cm}
\caption{\it The $(A_0/m_0, \tb)$ planes in the VCMSSM (left) and mSUGRA
  (right), without (upper) and with (lower panels) the
  \Och\ constraint~\protect\cite{WMAP}, showing the best-fit points and
  the 68 and 95\% CL contours in each case. 
}
\label{fig:A0tanb}
\end{figure*}

Finally, Fig.~\ref{fig:A0tanb} displays the $(A_0/m_0, \tb)$
planes for the VCMSSM (left) and mSUGRA (right). In the VCMSSM,
a wide range of values extending from $A_0/m_0 \sim +2$ down to
$\sim -3$ (corresponding to the case of very small $m_0$) 
lies within the 95\% CL contour,
whereas in mSUGRA only positive values are favoured at the 95\% CL
before the \Och\
constraint is applied. On the other hand, when it is applied, large
negative
values of $A_0/m_0$ are disfavoured also in the VCMSSM, and values of
$A_0/m_0 \sim 2$ with $\tb \sim 30$ are allowed at the 95\% CL.
Two distinct populations of preferred points are apparent in mSUGRA
when the \Och\ constraint is applied. There is a vestigial coannihilation
region with $A_0/m_0 \sim 2$ and $\tb \sim 28$, and the 
rapid-annihilation
funnel region has $A_0/m_0 \sim 0.4$ and $\tb \sim 7$.

We recall that the simplest Polonyi model of SUSY breaking~\cite{Polonyi}
predicts that $| A_0/m_0| = 3 - \sqrt{3}$, a possibility that is quite
consistent with the VCMSSM both before and after applying the \Och\
constraint  (left panels), but only marginally consistent with the 95\%
CL region of mSUGRA. 

\section{Comparison of Likelihood Analyses}

We gather in Table~\ref{tab:comparepoints} some important aspects
of the best-fit points in mSUGRA, the VCMSSM, the CMSSM and the NUHM1
(the latter being adapted from \cite{mc2,mc3}, with the inclusion of
the updated values of \gmt\ and $\mt$: we find that the values of
the different parameters at the best fit points in the CMSSM and the
NUHM1, as well as the fit probabilities, remain essentially unchanged 
with respect to the analysis in \cite{mc2,mc3}).
In the case of mSUGRA,
we list the properties of two local minima of the $\chi^2$
function: one is in the coannihilation region and one in the light Higgs
rapid-annihilation funnel with $\Delta \chi^2 \sim 4$, 
as discussed earlier.
We see in the first column that the mSUGRA fits have substantially higher
$\chi^2$ than the other models, which is reflected in the second column
by a significantly lower probability. The CMSSM and NUHM1 gave 
comparable fit probabilities of 32 (31)\% and the VCMSSM fit has a
probability of 31\%, while mSUGRA, with a probability of 6.0\% in
the coannihilation region and 
2.3\% in the light Higgs funnel region, provides a worse  
description of the data considered in this analysis. 
As a result, the mSUGRA scenario is somewhat disfavoured 
compared to the other SUSY scenarios we consider.
The source of tension in mSUGRA
is seen by comparing the third and fourth columns. The best
VCMSSM fit has very similar values of $m_{1/2}$ and $m_0$ to those in
the CMSSM, and in both cases $m_0 \ll m_{1/2}$. The conflict between
this preference and the cosmological requirement on mSUGRA that
$m_{3/2} = m_0 > \mneu{1}$ leads to a best fit in the coannhilation
region with larger values of $m_{1/2}$  and $\tb$ as well as a larger
value of $m_0$, and the other local minimum with small $m_{1/2}$ and
much larger $m_0$~%
\footnote{On the other hand, the best
NUHM1 fit has a value of $m_{1/2}$ similar to those in the VCMSSM and
CMSSM but a somewhat larger value of $m_0$, which is possible because an
acceptable value of \Och\ may be found along a funnel in the $(m_0,
m_{1/2})$ plane at lower values of $\tb$, due to $\neu{1} \neu{1}$
annihilation through direct-channel $H, A$ poles. This option is not
available in the CMSSM except at large $\tb$ and $m_{1/2}$, and it is
absent completely in mSUGRA.}%
.~As seen in the sixth column,
the values of $\tb$ favoured in the VCMSSM, CMSSM and 
NUHM1 are much smaller than that favoured 
at the best-fit point of mSUGRA in the coannihilation region. 
The last column shows the best-fit values of $\Mh$, {\em not} taking
into account the LEP and Tevatron limits. It can be seen
that the best-fit values of $\Mh$ in mSUGRA in the coannihilation
region, the VCMSSM and the CMSSM are similar, and  somewhat below the
LEP lower limit, leading to increases in $\chi^2$ of 3.9 (1.1) (1.4)
when the LEP constraint is applied. On the other hand, we find rather
higher best-fit values of $\Mh$ in the NUHM1 and mSUGRA in the funnel
region. Finally, we note that the largest variations between the
different models occur for $A_0$ (fifth column), 
reflecting the relative insensitivity of our fits to this parameter.

\begin{table*}[!tbh]
\renewcommand{\arraystretch}{1.2}
\begin{center}
\begin{tabular}{|c||c|c|c|c|c|c||c|c|} \hline
Model & Minimum $\chi^2$ & Probability & $m_{1/2}$ & $m_0$ & $A_0$ & 
$\tb$ & $\Mh$ (no LEP) \\ \hline \hline
mSUGRA & 29.4 & 6.0\% & 550 & 230  & 430 & 28 & 107.7 \\
       & 33.2 & 2.3\% & 130 & 2110 & 980 & 7  & 116.9 \\
VCMSSM & 22.5 & 31\% & 300 & 60 &  30 & 9 & 109.3 \\
CMSSM & 21.3 & 32\% & 320 & 60 & -160 & 11 & 107.9 \\
NUHM1 & 19.3 & 31\% & 260 & 100 & 1010 & 8 & 119.5 \\
\hline
\end{tabular}
\caption{\it Comparison of the best fits found in this paper within the
  mSUGRA and VCMSSM frameworks with previous results~\cite{mc2,mc3}
in the CMSSM and NUHM1 frameworks. In addition to the minimum value of
$\chi^2$ in each scenario, we include the values of $m_{1/2}, m_0, A_0$
and $\tb$ at  all the best-fit points, as well as $\Mh$ 
(for the latter the direct bounds from LEP and the Tevatron are {\em not}
included). All masses are in GeV units. We list two
very different mSUGRA fit results 
with similar $\chi^2$ values: the first is in
the coannihilation region, and the second is in the light Higgs funnel
region. Note that we use here the convention for $A_0$ described in the
text, which differs from that in~\cite{mc1}.
\label{tab:comparepoints}} 
\end{center}
\end{table*}

\begin{figure*}[htb!]
\begin{center}
\resizebox{10cm}{!}{\includegraphics{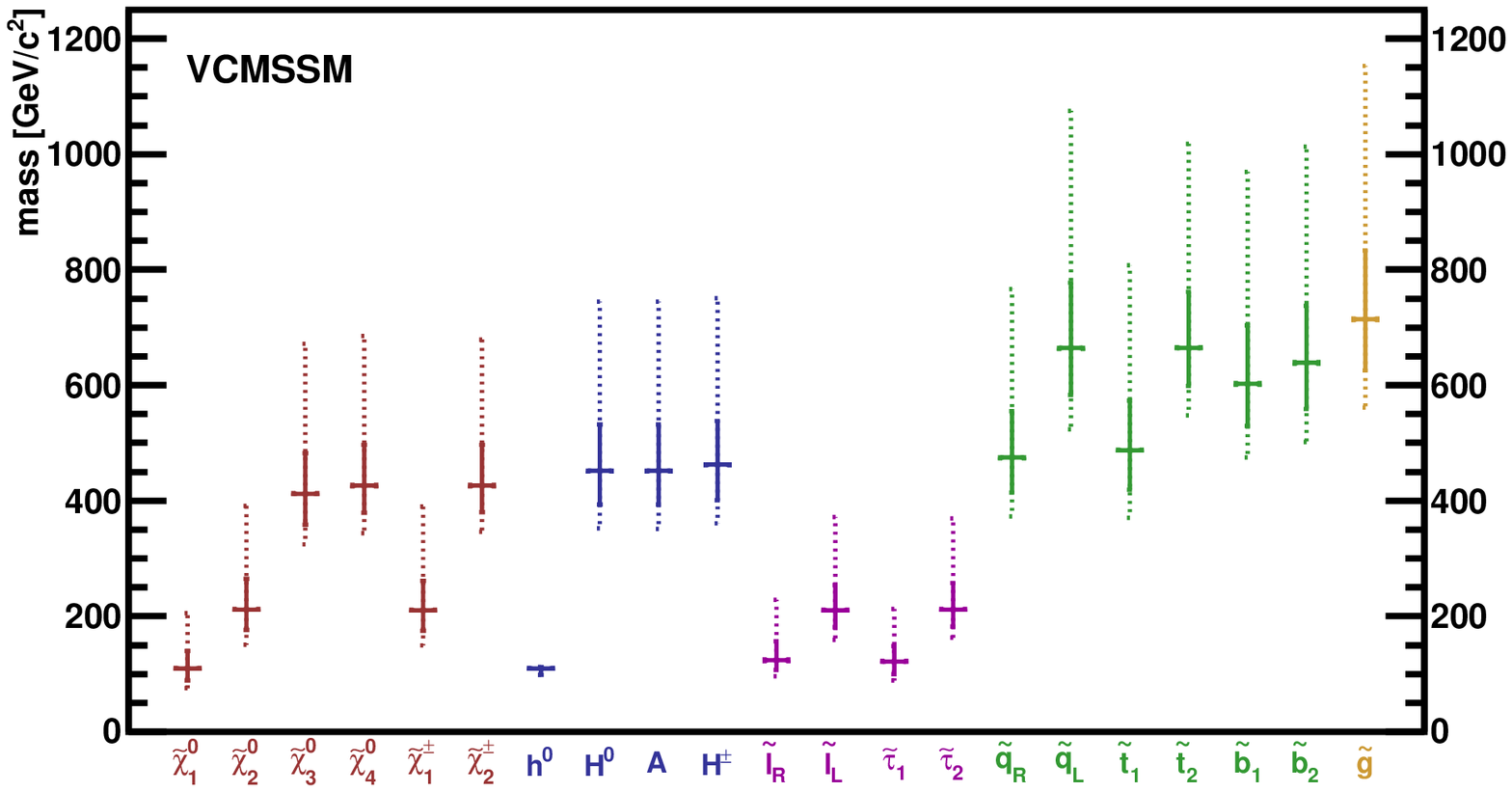}} \\
\resizebox{10cm}{!}{\includegraphics{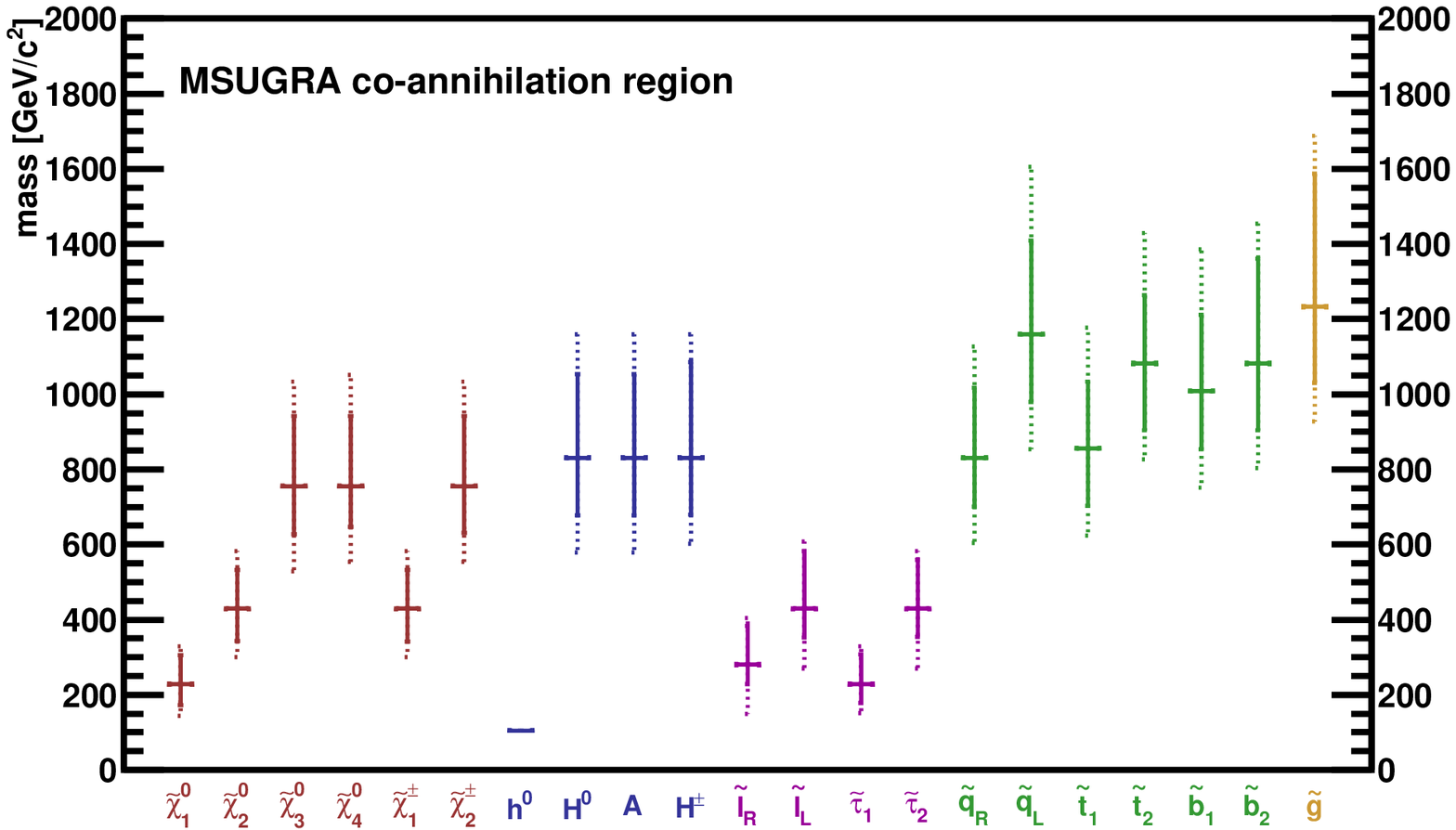}} \\
\resizebox{10cm}{!}{\includegraphics{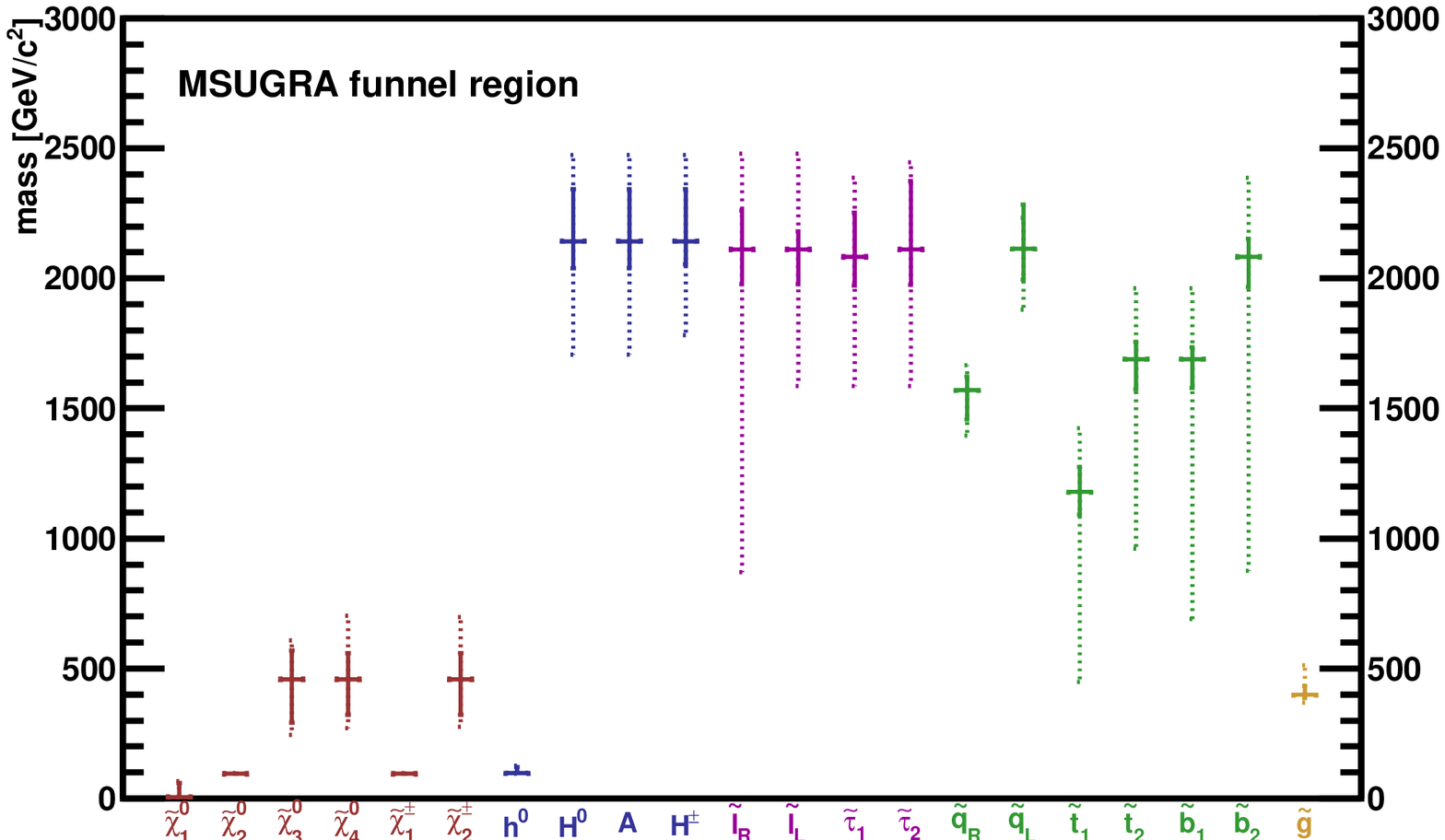}} \\
\end{center}
\caption{\it Spectra in the VCMSSM (top), and mSUGRA in the
coannihilation region (middle) and the funnel region (bottom),
  implementing all the constraints including that on \Och. The horizontal
solid lines indicate the best-fit values, the vertical solid lines
are the 68\% CL ranges, and the vertical dashed lines are the 95\% CL
ranges for the indicated mass parameters. 
}
\label{fig:spectrum}
\end{figure*}

In Fig.~\ref{fig:spectrum} we display the spectra in the VCMSSM (top)
and mSUGRA (middle and bottom) with the \Och\ constraint applied,
complementing the CMSSM and NUHM1 spectra shown in Fig.~3
of~\cite{mc3}. In the case of the VCMSSM, the spectrum is
qualitatively similar to those in the CMSSM and NUHM1~\cite{mc2,mc3}.
The two mSUGRA spectra are in the coannihilation (middle) and funnel
region (bottom), reflecting the coexistence of two qualitatively
different (near-)best-fit points with relatively similar $\chi^2$. The
spectra in these regions are significantly different from each other and
from the VCMSSM, CMSSM and NUHM1. This is because the coannihilation
region has $m_{1/2}$ significantly larger than in the 
other models, whereas the funnel region has a significantly smaller and very
well-defined value of $m_{1/2}$ and relatively large values of
$m_0$. This bimodality 
affects directly the preferred values of $\mneu{1}$ and $\mgl$, and affects the
other sparticle masses via renormalization effects.
These spectra show that the coloured particles are within
the reach of the LHC for the VCMSSM and mSUGRA  in the coannihiliation
region, whereas more integrated luminosity would be
necessary for mSUGRA in the funnel region (except for gluino production). 
In each scenario some SUSY
particles should be accessible at an $e^+e^-$ collider, even with a
center-of-mass energy as low as $500 \gev$.

\begin{figure*}[htb!]
\begin{center}
\resizebox{7.9cm}{!}{\includegraphics{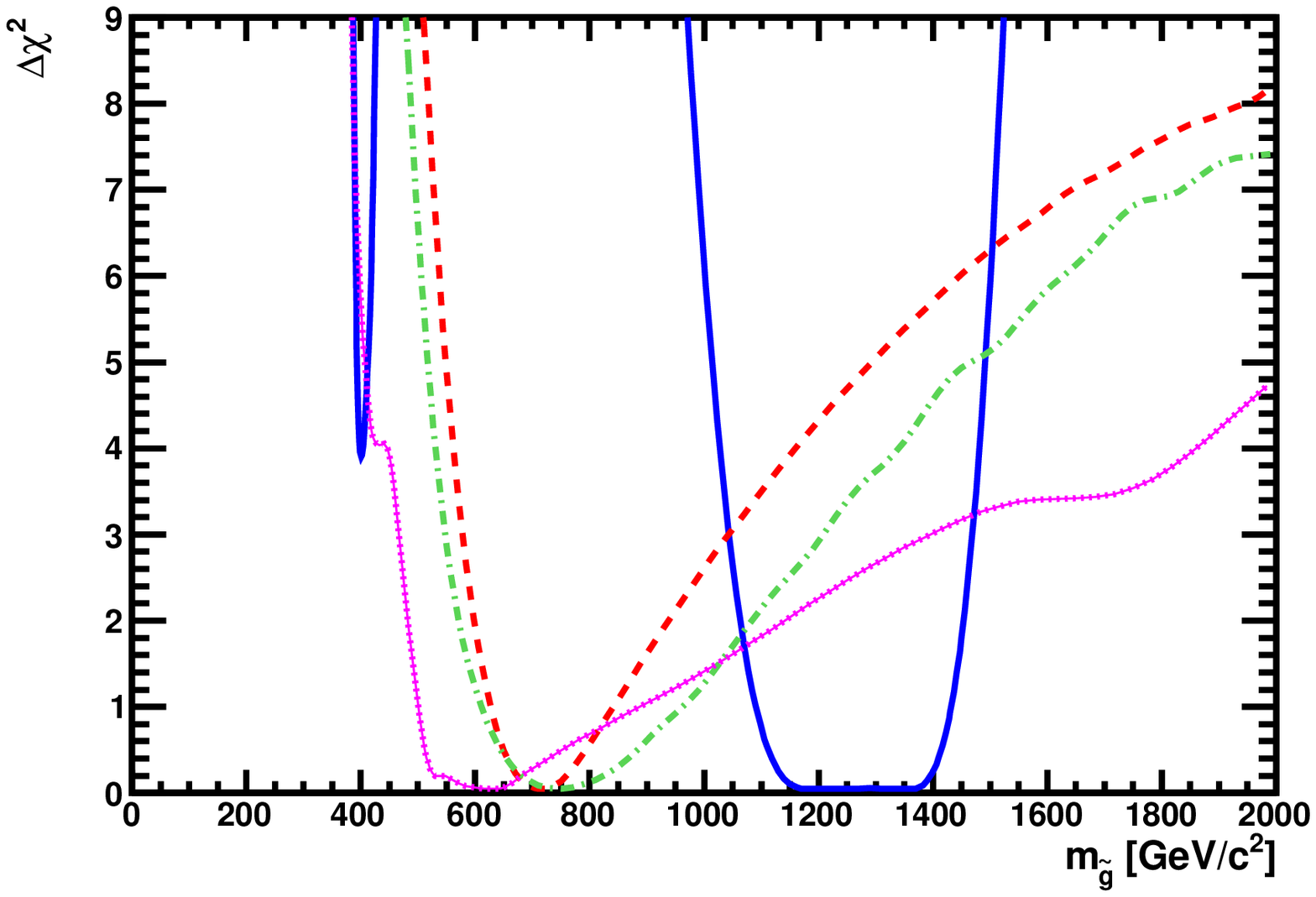}}
\resizebox{7.9cm}{!}{\includegraphics{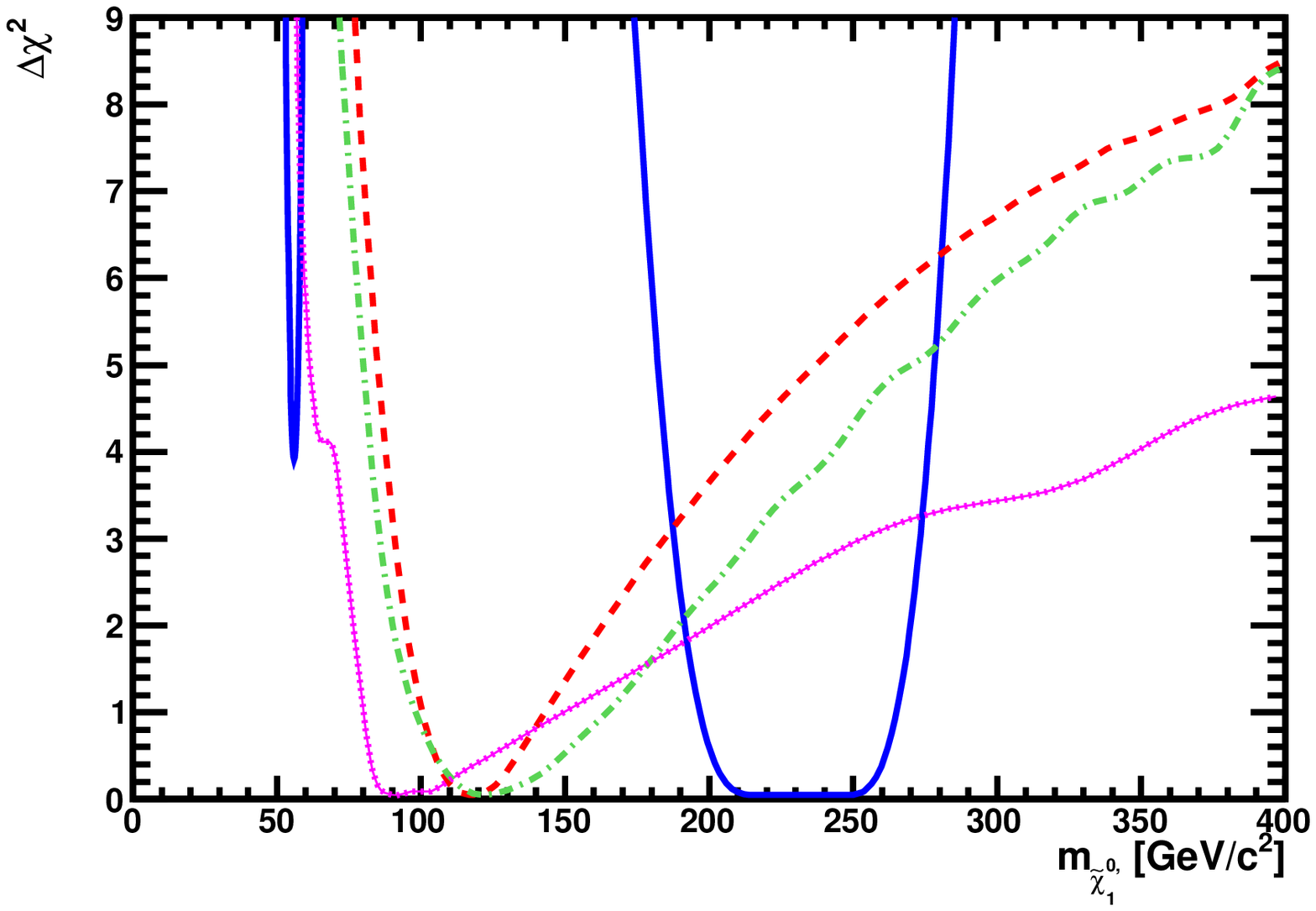}} \\[0.5em]
\resizebox{7.9cm}{!}{\includegraphics{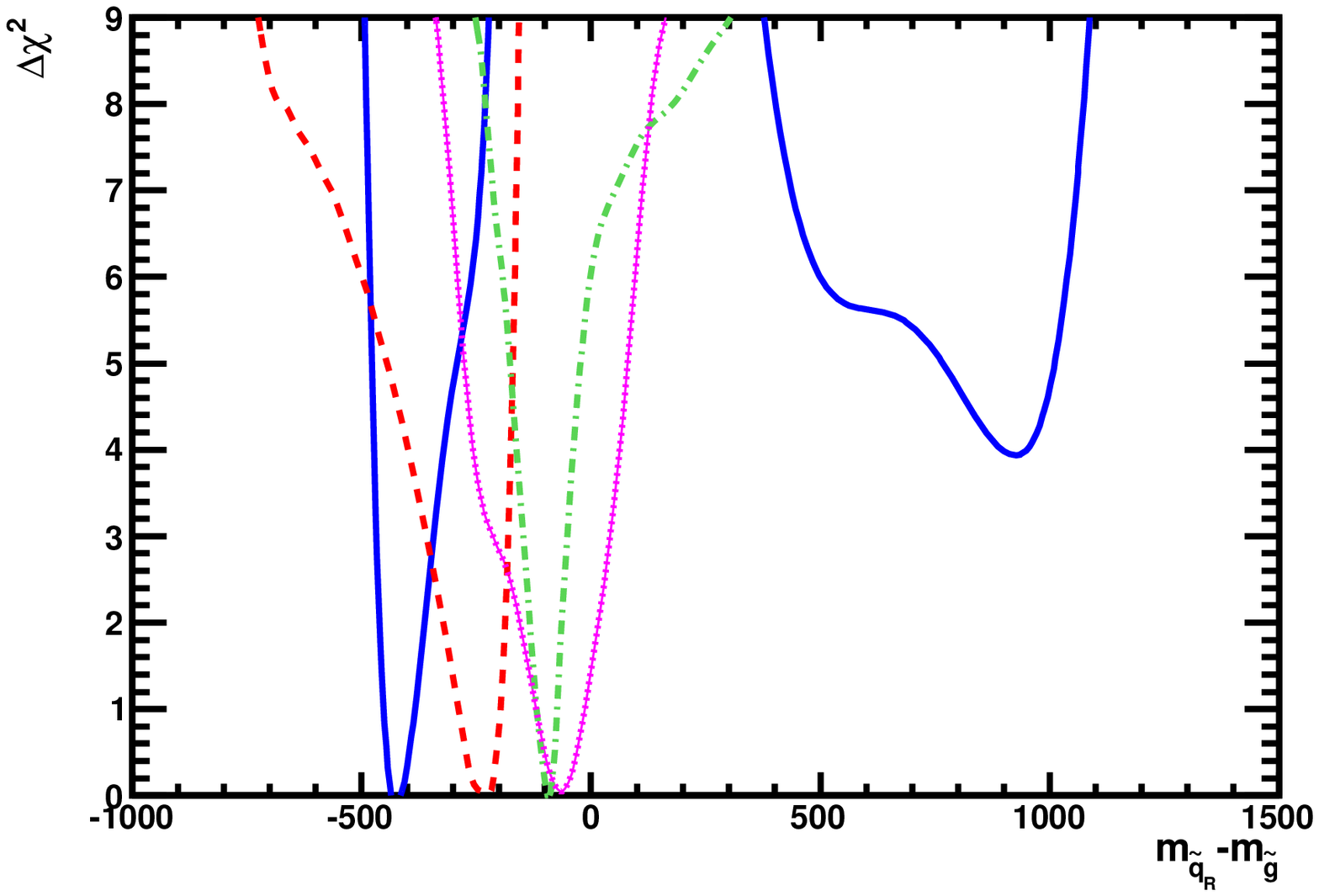}}
\resizebox{7.9cm}{!}{\includegraphics{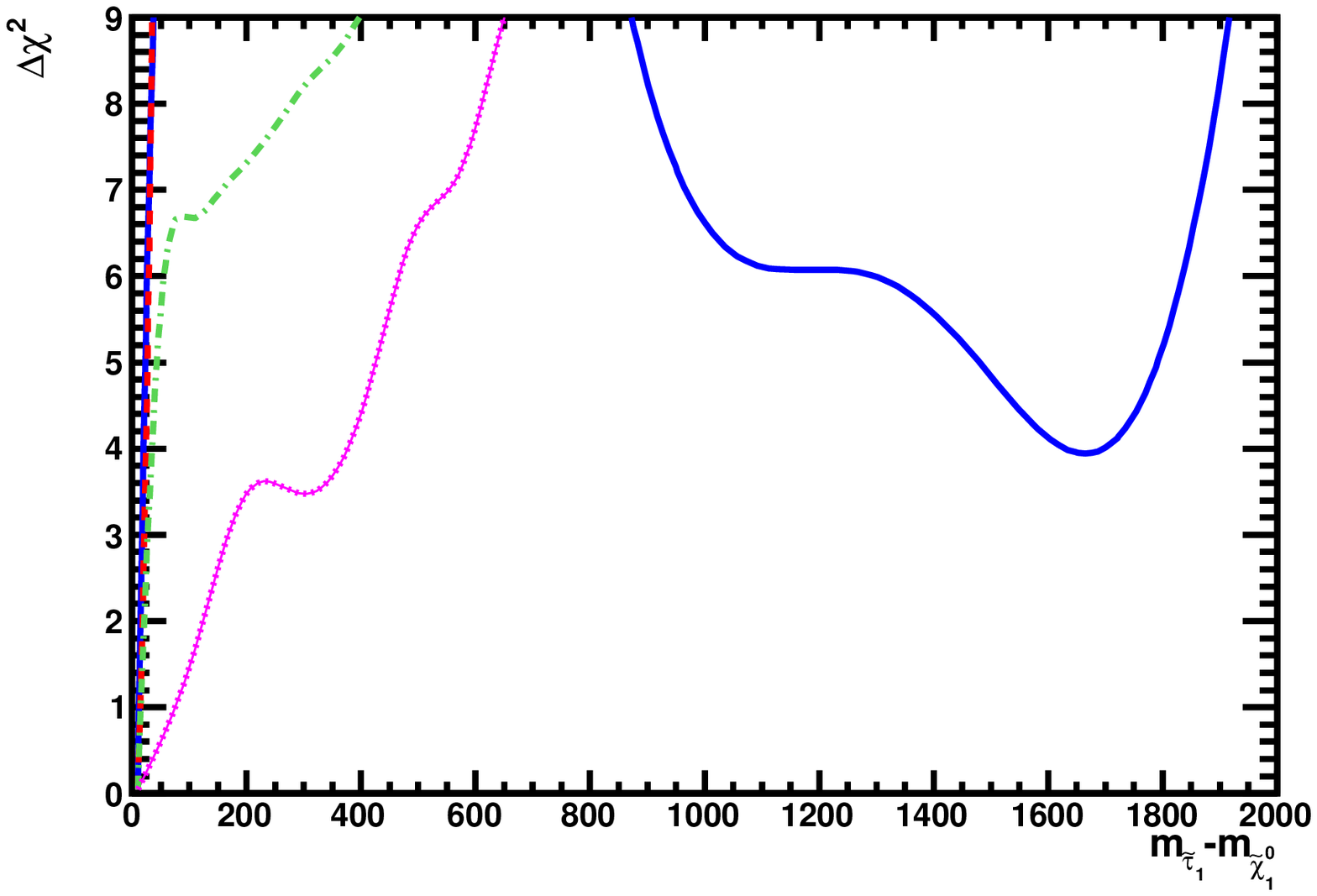}} \\[0.5em]
\resizebox{7.9cm}{!}{\includegraphics{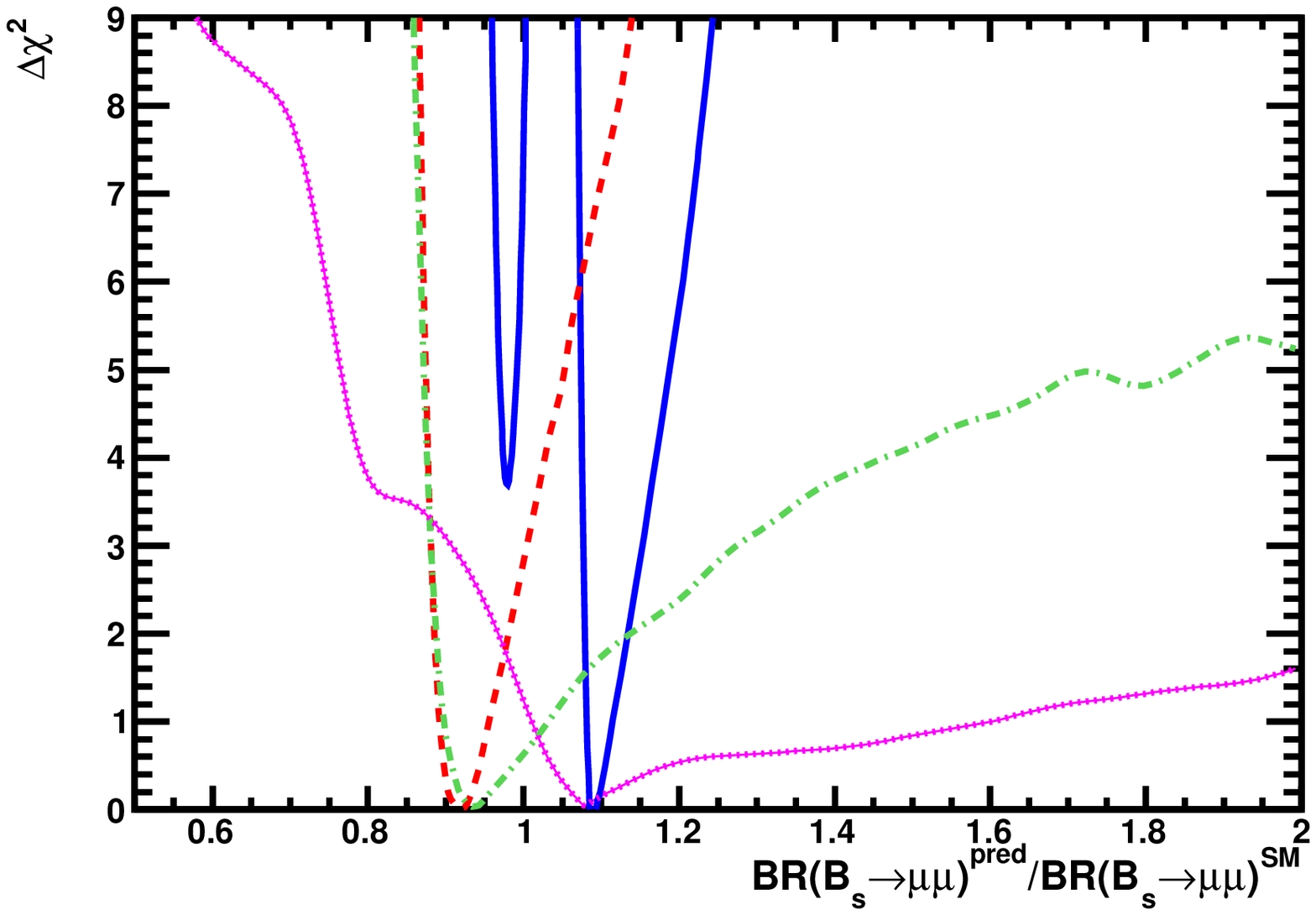}}
\resizebox{7.9cm}{!}{\includegraphics{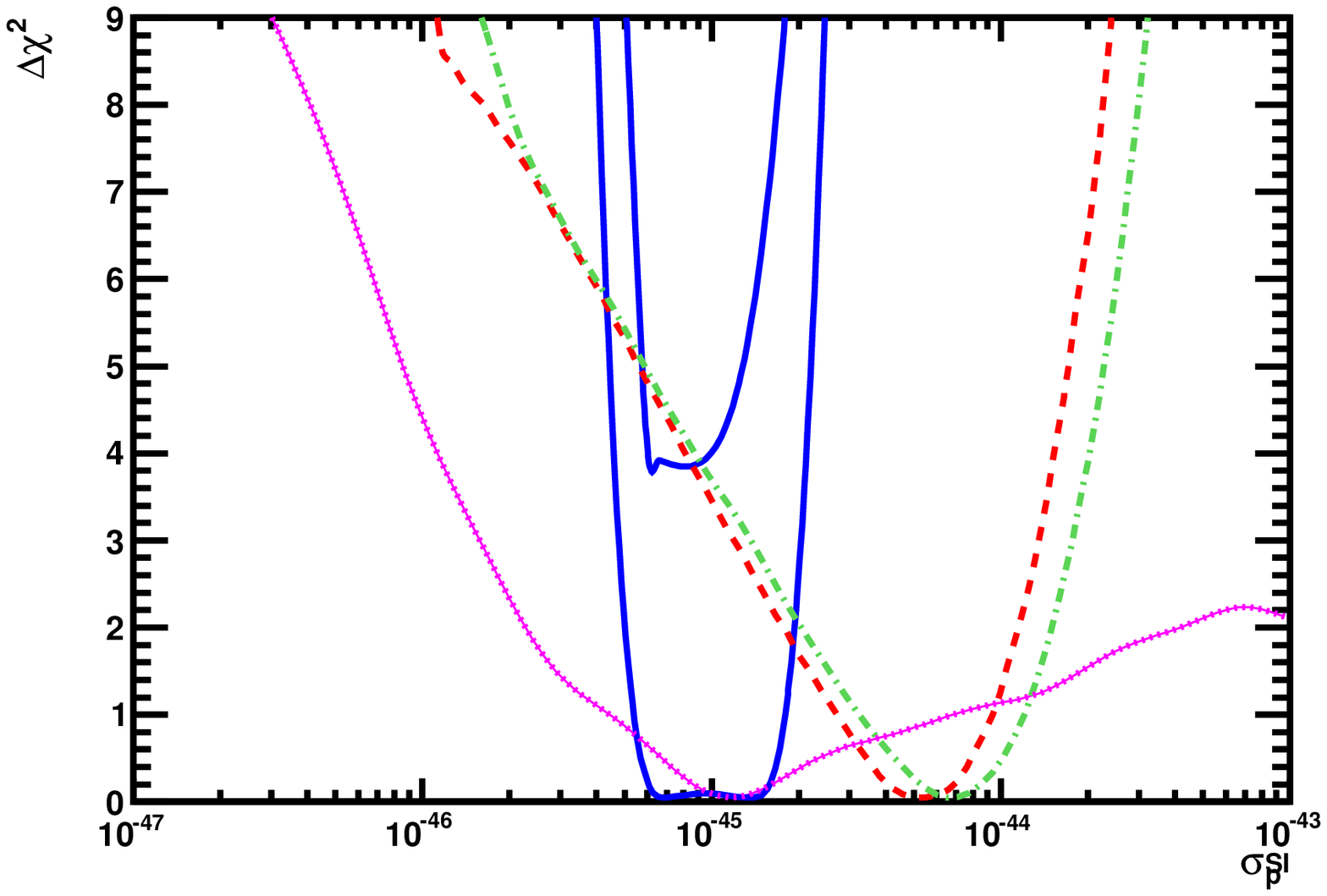}} 
\end{center}
\caption{\it The likelihood functions for (top) $m_{\tilde g}$ and $\mneu{1}$, 
(middle) $m_{\tilde q_R} - m_{\tilde g}$ and $\mstaue - \mneu{1}$, and 
(bottom) \bmm\ (normalized to the SM estimate) and $\ssi$. In each case,
we compare the predictions in mSUGRA (blue solid line), 
the VCMSSM (red dashed line), 
the CMSSM (green dash-dotted line)
and the NUHM1 (purple dotted line).
}
\label{fig:1Dlikelihoods}
\end{figure*}

We display in Fig.~\ref{fig:1Dlikelihoods} the one-dimensional $\chi^2$
likelihood functions for (top panels) $\mgl$ and $\mneu{1}$, (middle
panels) the mass differences $m_{\tilde q_R} - \mgl$ and 
$\mstaue - \mneu{1}$, and (bottom panels) \bmm\ and $\ssi$, the
spin-independent neutralino scattering cross section. 
The NUHM1 curves are shown as purple dotted lines, the CMSSM curves as
green dash-dotted lines, the VCMSSM curves as red dashed lines, and the
mSUGRA curves as 
blue solid lines. In each plot, we display the $\Delta\chi^2$ contribution
of each model relative to the best-fit point in that model. Thus
the secondary minimum of the $\chi^2$ function for mSUGRA
has the appropriate $\Delta \chi^2$ relative to the absolute minimum in
that model. 

We see that the likelihood functions for $\mgl$ and $\mneu{1}$ in the
VCMSSM are similar to those in the CMSSM and NUHM1~\cite{mc2,mc3},
with the most likely range of $\mgl \sim 700$ to $800 \gev$ and 
$\mneu{1} \sim 120 \gev$. 
The corresponding likelihood functions in mSUGRA, on the other hand, 
are very different, reflecting once more the bimodality in $m_{1/2}$ and $m_0$.
In the VCMSSM, as in the CMSSM and NUHM1, the preferred range of
$\mgl$ (top left panel) suggests that there may be good
prospects for observing first hints of SUSY at the LHC in 2011/12, 
although values of $\mgl > 2000 \gev$ are permitted in the NUHM1
  with $\Delta\chi^2 \sim 5$.
In the rapid-annihilation strip of mSUGRA, $\mgl \sim 400 \gev$ and
$\mneu{1} \sim 55 \gev$, putting the discovery of the gluino at the LHC
within the reach of the LHC in 2011/12, and making $\mgl$ potentially
a powerful diagnostic tool for mSUGRA. 
However, discovery of the 
gluino would be delayed in the (more probable, but still
disfavoured scenario compared to VCMSSM, CMSSM and NUHM1)
coannihilation region of mSUGRA, where the best fit has 
a gluino mass around 1200 to $1400 \gev$.

The bimodality of the mSUGRA likelihood function is reflected again in the 
predictions for $m_{\tilde q_R} - \mgl$ (middle left) and $\mstaue - \mneu{1}$
(middle right) in Fig.~\ref{fig:1Dlikelihoods}. 
The two different mSUGRA minima yield different signs
for $m_{\tilde q_R} - \mgl$, whereas $\mgl > m_{\tilde q_R}$ is favoured
in the VCMSSM, as was shown previously to be favoured in the
CMSSM~\cite{mc2,mc3}, and to a rather lesser extent in the
NUHM1~\cite{mc2,mc3}. Thus, the 
sign and magnitude of $m_{\tilde q_R} - \mgl$ are potential diagnostic
tools for discriminating between different models~\footnote{We recall that
  dijet + missing energy events due to the decay chain 
${\tilde g} \to {\tilde q_R} + {\bar q}, {\tilde q_R} \to q + \neu{1}$ 
are in general favoured when $m_{\tilde q_R} < \mgl$,
whereas there are expected to be a larger fraction of four-jet + missing
energy events when $m_{\tilde q_R} > \mgl$, and different decay chains occur via
${\tilde q_L}$, ${\tilde t_{1,2}}$ and ${\tilde b_{1,2}}$.}.
We note also the different predictions for $\mstaue - \mneu{1}$: the
VCMSSM predicts a very small mass difference, as was shown previously
in the CMSSM, and so does mSUGRA in the coannihilation region. However, in the
rapid-annihilation region mSUGRA predicts mass
differences that may be large, as was previously shown to be possible 
(to a lesser extent) in the NUHM1~\cite{mc3}. 
Thus discovery of a light gluino at the LHC and/or a light
neutralino LSP would not necessarily imply that the lightest 
slepton would have a mass close to that of the LSP,
and the lighter stau could be too heavy for an
$e^+e^-$ collider with $\sqrt{s} = 1 \tev$.
However, in this scenario the whole chargino and neutralino
spectrum would be accessible at an $e^+e^-$ collider with 
$\sqrt{s} = 1 \tev$, see Fig.~\ref{fig:spectrum}.

We see in the bottom left panel of Fig.~\ref{fig:1Dlikelihoods} that
in the VCMSSM a value of  \bmm\ slightly lower than the SM value 
is favoured, although larger values cannot be excluded. 
On the other hand, in mSUGRA in the coannihilation region, 
with its relatively large values of $\tb$, we find a preferred value of 
\bmm\ that is
slightly larger than in the SM, whereas a range around the SM level
is favoured in the light Higgs funnel region.
As already seen in~\cite{mc3}, the minimum of $\chi^2$ in the NUHM1 is
at a somewhat higher value of \bmm\ than in the SM, whereas the
best-fit value in the CMSSM is again slightly smaller, albeit 
with considerable uncertainty. The values below the SM prediction arise from
chargino-induced $Zbs$ penguin diagrams, that yield analogous but smaller
reductions in BR($K \to \pi \nu {\bar \nu}$). However, the SM prediction
has an uncertainty of about 10\%, and ${\cal O}(100)$ $B_s \to \mu^+ \mu^-$ 
decays would be needed to match this error, so the differences between
the mSUGRA, VCMSSM and best-fit 
CMSSM predictions are probably unobservable.
 
Turning to the likelihood functions for $\ssi$ in the bottom right panel
of Fig.~\ref{fig:1Dlikelihoods} (calculated assuming a $\pi$-N scattering
$\sigma$ term
$\Sigma_N = 64 \mev$: see, e.g., \cite{Savage} for a discussion of the
implications of modifying this assumption) we see that a range between
$10^{-45}$ and $10^{-44}$~cm$^2$ is favoured in the VCMSSM, whereas in mSUGRA
a range between $2 \times 10^{-46}$~cm$^2$ and 
$5 \times 10^{-45}$~cm$^2$ is favoured.

\begin{figure*}[htb!]
\resizebox{8cm}{!}{\includegraphics{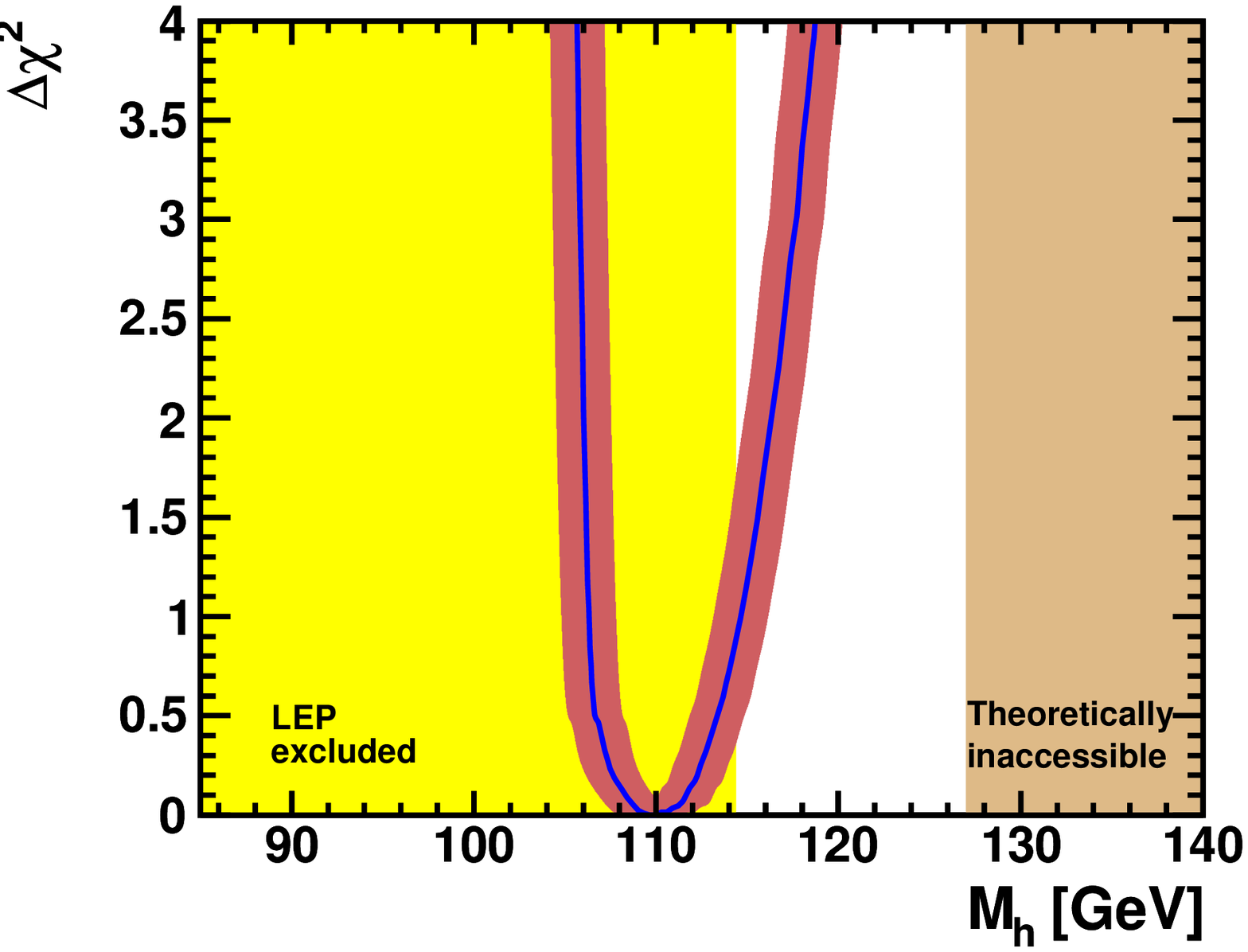}}
\resizebox{8cm}{!}{\includegraphics{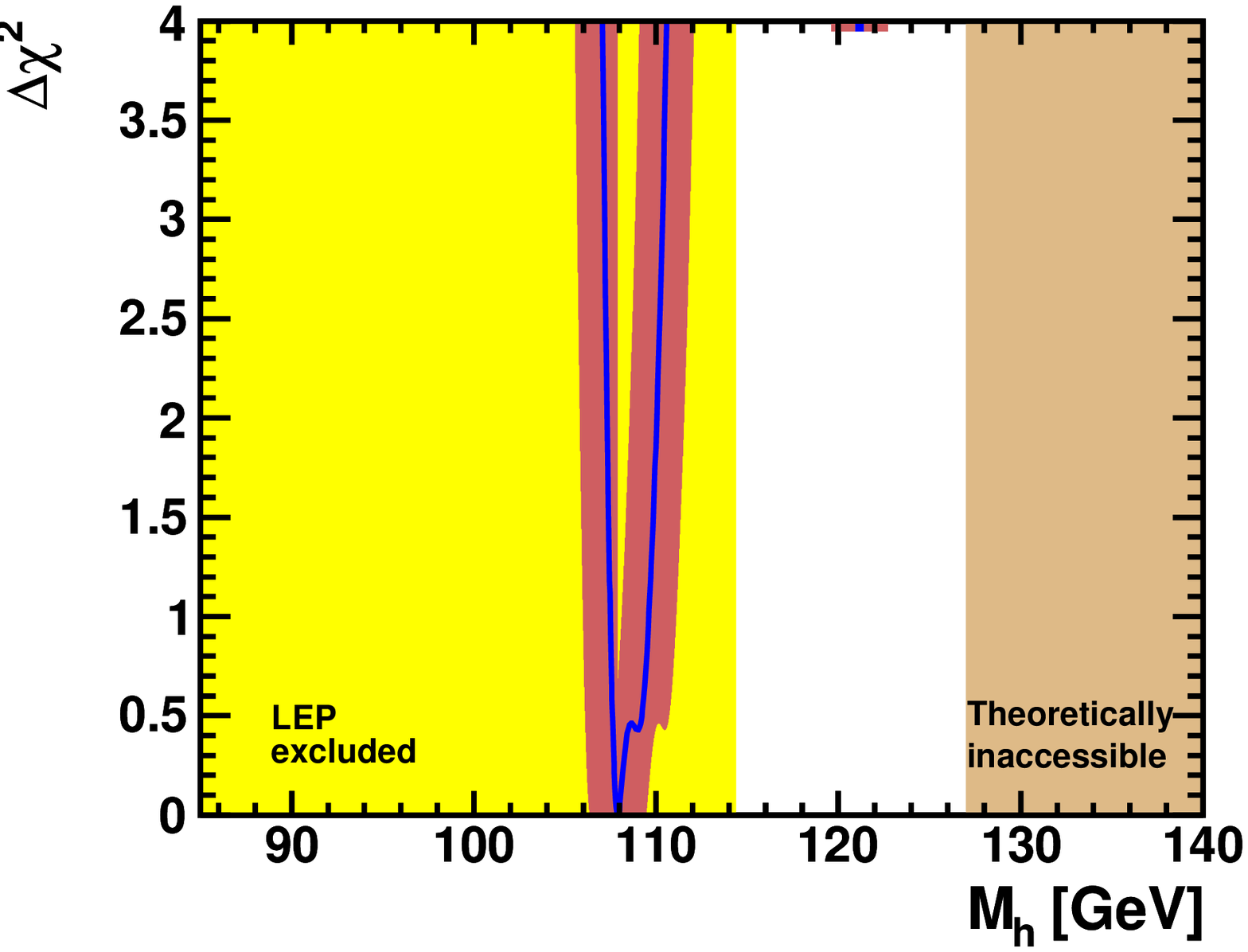}}
\caption{\it The $\chi^2$ functions for $\Mh$ in the VCMSSM (left) and
  mSUGRA (right), including the theoretical uncertainties (red
  bands), as well as the SM prediction for $\Mh$ based on a
  precision electroweak fit (blue band). 
Also shown is the mass range for a SM-like Higgs
  boson excluded by LEP (yellow shading), and the mass range
  that is theoretically inaccessible in TeV-scale SUSY (beige shading).
Note in right panel the secondary minimum at 
$\Mh = 121.1 \gev, \Delta \chi^2 = 3.9$ 
and the accompanying narrow horizontal red band.
}
\label{fig:redband}
\end{figure*}

Finally, in Fig.~\ref{fig:redband} we display the one-dimensional
$\chi^2$ likelihood functions for $\Mh$ in the VCMSSM (left) and mSUGRA
(right), {\em not} including the direct limits from
LEP and the Tevatron~\footnote{See~\cite{mc1,mc2,mc3} for the
corresponding plots in the CMSSM and NUHM1.}. 
For each model we display the likelihood functions, 
including the theoretical uncertainties (red bands),
which we take to be $1.5 \gev$ in all models. For
comparison, we also show the mass range excluded for a SM-like Higgs
boson (yellow shading) obtained at
LEP~\cite{Barate:2003sz,Schael:2006cr}. This limit is valid since the
VCMSSM and mSUGRA are sub-spaces of the more general CMSSM parameter
space, where the LEP limits have been shown to be
valid~\cite{Ellis:2001qv,Ambrosanio:2001xb}. 
Values somewhat below the LEP exclusion are favoured in the
coannihilation region of mSUGRA, in the VCMSSM and the CMSSM, entailing
$\chi^2$ prices of 3.9 (1.1) (1.4), as discussed previously. In the
mSUGRA coannihilation case, the global minimum of $\chi^2$ found when
the LEP constraint is disregarded is in an isolated region at low $(m_0,
m_{1/2})$. When the LEP Higgs constraint is applied, this region is
strongly disfavoured, and the global minimum moves to the green star
shown in Fig.~\ref{fig:m0m12}, located at much larger $(m_0, m_{1/2})$
and with $\Delta \chi^2 = 3.9$. This other minimum is reflected in the
blue point at $\Mh = 121.1 \gev$ with $\Delta \chi^2 = 3.9$ and the
corresponding horizontal red line visible at the top of the right panel
in Fig.~\ref{fig:redband}. Comparably large values of  $\Mh \sim 117$,
$120 \gev$ are favoured in the funnel region of mSUGRA and in the
NUHM1, so these models are naturally consistent with the LEP bound on a
SM-like Higgs boson. 

\section{Conclusions}

We have completed in this paper the frequentist analysis of a nested
sequence of variants of the MSSM: NUHM1 $\ni$ CMSSM $\ni$ VCMSSM $\ni$
mSUGRA, discussing in each case the best-fit point, the minimum
$\chi^2$/dof, the 68 and 95\% CL regions and aspects of the favoured
ranges of particle masses and other observables. We found previously
that the restriction from the NUHM1 to the CMSSM does not change
drastically the position of the best-fit point or the favoured ranges of
parameters and observables. This reflects the fact that the present data
do not constrain significantly the heavier Higgs bosons of the MSSM, and so
there is no significant tension in the CMSSM fit arising from that
sector. Likewise, we found in the present
paper that the restriction from the
CMSSM to the VCMSSM  does not have a large impact on the position of the
best-fit point or on the favoured ranges. This reflects the fact that
the present data also do not constrain significantly the trilinear soft
SUSY-breaking parameter $A_0$, and so there is no significant $\chi^2$
price to be paid when setting $A_0 = B_0 + m_0$ as in the VCMSSM. Fits
in all three of the NUHM1, CMSSM and VCMSSM frameworks have good
absolute probabilities in our frequentist analyses. 

On the other hand, in mSUGRA two almost equally good best-fit points
coexist, with rather different MSSM parameter values, significantly
higher values of $\chi^2$ and lower absolute
probability, somewhat disfavouring this scenario. 
This reflects the fact that the neutralino LSP constraint
$\mneu{1} < m_{3/2}$ excludes the best fit found in the VCMSSM and forces
instead either rather larger values of $m_0, m_{1/2}$ and $\tb$ or
points along the narrow light Higgs rapid-annihilation funnel with small
$m_{1/2}$ but large $m_0$.  

In each of the NUHM1, CMSSM and VCMSSM there is a significant chance of
observing first hints of SUSY in the 2011/12 run of the LHC, whereas 
this may be 
more problematic in mSUGRA, depending on whether its parameters lie in the
rapid-annihilation funnel or in the higher-mass coannihilation
region. Correspondingly, early hints of SUSY at the LHC in this first run
might further favour the NUHM1, CMSSM and VCMSSM frameworks over the
higher-mass mSUGRA option in the coannihilation region. In the
rapid-annihilation funnel region of mSUGRA one would expect abundant
production of gluinos, whereas the first appearances of squarks, having
masses in excess of $1 \tev$, would only happen at a somewhat later
stage. If no sign
of SUSY particles shows up in the early LHC run, in particular if the 
integrated luminosity (and energy) goes significantly beyond the
currently foreseen 1/fb at $7 \tev$, the scenarios predicting a
relatively low SUSY scale could soon come under pressure.
At the time of writing, the ATLAS
and CMS collaborations are each examining some 35/pb of 
analysable data.

\subsection*{Note added in proof}

Since the submission of this paper, the first results of searches for
supersymmetry have been published by the CMS and ATLAS
Collaborations~\cite{CMSsusy,ATLASsusy}. These change somewhat the
parameters of the best-fit points and the 68\% and 95\% CL regions in
the models studied here, but do not disfavour any of them
strongly~\cite{mc5}. 


\subsection*{Acknowledgements}

We thank Gino Isidori for valuable discussions.
Work supported in part by the European Community's Marie-Curie Research
Training Network under contract MRTN-CT-2006-035505
`Tools and Precision Calculations for Physics Discoveries at Colliders'. 
The work of KAO was supported in part by
DOE grant DE-FG02-94ER-40823 at the University of Minnesota.


\end{document}